\documentclass[a4paper,12pt]{article}

\usepackage{jcappub} 
\usepackage{hyperref}
\hypersetup{
    bookmarks=true,         
    unicode=false,          
    pdftoolbar=true,        
    pdfmenubar=true,        
    pdffitwindow=false,     
    pdfstartview={FitH},    
    pdftitle={My title},    
    pdfauthor={Author},     
    pdfsubject={Subject},   
    pdfcreator={Creator},   
    pdfproducer={Producer}, 
    pdfkeywords={keyword1} {key2} {key3}, 
    pdfnewwindow=true,      
    colorlinks=true,       
    linkcolor=red,          
    citecolor=cyan,        
    filecolor=magenta,      
    urlcolor=blue,           
    linktocpage=true
}
\usepackage{subfigure}
\usepackage{pstricks}
\usepackage{color}
\usepackage{amsfonts}
\usepackage{mathrsfs}
\usepackage{epsfig}
\usepackage{amsmath,amssymb,amsthm,graphicx,latexsym}
\usepackage{ulem}

\newcommand{\be}{\begin{equation}}
\newcommand{\ee}{\end{equation}}
\newcommand{\bea}{\begin{eqnarray}}
\newcommand{\eea}{\end{eqnarray}}
\newcommand{\bml}{\begin{subequations}}
\newcommand{\eml}{\end{subequations}}
\newcommand{\bfig}{\begin{figure}}
\newcommand{\efig}{\end{figure}}

\newcommand{\slashed}{\hspace{-1.1ex}/}

\title{
  Constraining ${\cal N}=1$ supergravity inflation with non-minimal K\"ahler operators  
 using $\delta N$ formalism }

\author[a]{ Sayantan Choudhury}

\affiliation[a]{ Physics and Applied Mathematics Unit, Indian Statistical Institute, 203 B.T. Road, Kolkata 700 108, INDIA}

\abstract{In this paper I provide a general framework based on $\delta N$ formalism to study the features
 of unavoidable higher dimensional non-renormalizable K\"ahler operators for ${\cal N}=1$ supergravity (SUGRA)
during primordial inflation from the combined constraint on non-Gaussianity, sound speed and CMB dipolar asymmetry as obtained from the recent Planck data.
In particular I study the nonlinear evolution of cosmological perturbations on
large scales which enables us to compute the curvature perturbation, $\zeta$, 
without solving the exact perturbed field equations.
Further I compute the non-Gaussian parameters $f_{NL}$ , $\tau_{NL}$ and $g_{NL}$
for local type of non-Gaussianities and CMB dipolar asymmetry parameter, $ A_{CMB}$, using the $\delta N$ formalism 
for a generic class of sub-Planckian models induced by the Hubble-induced corrections for a 
minimal supersymmetric D-flat direction where inflation occurs at the point of inflection within the visible sector.
 Hence by using multi parameter scan I constrain the 
non-minimal couplings appearing in non-renormalizable K\"ahler operators within, ${\cal O}(1)$, for the
 speed of sound, $0.02\leq c_s\leq 1$, and tensor to scalar, $10^{-22} \leq r_{\star} \leq 0.12$.
Finally applying all of these constraints I will fix the lower as well as the upper bound of the non-Gaussian parameters within, ${\cal O}(1-5)\leq f_{NL}\leq 8.5$,
 ${\cal O}(75-150)\leq\tau_{NL}\leq 2800$ and ${\cal O}(17.4-34.7)\leq g_{NL}\leq 648.2$,
and CMB dipolar asymmetry parameter within the range, $0.05\leq A_{CMB}\leq 0.09$.}

\begin{document} 
\maketitle
\flushbottom

\section{Introduction}

The primordial inflationary paradigm is a very rich idea to explain various aspects of the early universe, which creates
 the perturbations and the matter. For recent developments see Refs.~\cite{Mazumdar:2010sa,Mazumdar:2011zd}. Usually inflation prefers slow-rolling of a single scalar field on
a flat potential, which has unique predictions for the Cosmic Microwave Background (CMB) observables. 
The induced cosmological perturbations are generically random Gaussian in nature with
a small tilt and running in the primordial spectrum which indicates that inflation must come to
an end in our patch of the universe. But a big issue may crop up in model discrimination and also in the removal of the degeneracy of cosmological
parameters obtained from CMB observations \cite{Spergel:2006hy,Ade:2013uln,Ade:2013zuv,Ade:2013ydc}.  
Non-Gaussianity has emerged as a powerful observational tool to discriminate between different
 classes of inflationary models \cite{Maldacena:2002vr,Bartolo:2004if,Komatsu:2009kd,Chen:2010xka,Komatsu:2010hc}.
 The Planck data show no significant evidence in favour of primordial non-Gaussianity, the current limits \cite{Ade:2013ydc} are
yet to achieve the high statistical accuracy expected from the single-field inflationary models and for this opportunities are galore
for the detection of large non-Gaussianity from various types of inflationary models.
To achieve this goal, apart from the huge success of cosmological linear perturbation
theory, the general focus of the theoretical physicists has now shifted towards the study of nonlinear evolution of
cosmological perturbations. Typically any types of
nonlinearities are expected to be small; but, that can be estimated via non-Gaussian n-point correlations of cosmological perturbations.
The so-called ``$\delta N$ formalism'' (where $N$ being the number of e-foldings)
 \cite{Starobinsky:1982,Salopek:1990,Sasaki:1995aw,Wands:2000dp,Lyth:2004gb,Lyth:2005fi,Mazumdar:2012jj,Sugiyama:2012tj} is a well accepted tool for computing non-linear evolution of cosmological
perturbations on large scales ($k<<aH$), which is derived using the ``separate universe''
 approach \cite{Sasaki:1995aw,Wands:2000dp,Sasaki:1998ug,Sasaki:2007ay}. Particularly, it provides a fruitful technique to
compute the expression for the curvature perturbation $\zeta$ without explicitly
solving the perturbed field equations from which the various
 local non-Gaussian parameters, $f_{NL}^{\mathrm{local}},\tau_{NL}^{\mathrm{local}},g_{NL}^{\mathrm{local}}$ and
 CMB dipolar asymmetry parameter \cite{Erickcek:2008sm,Lyth:2013vha,Wang:2013lda,Dai:2013kfa,Mazumdar:2013yta}, $A_{CMB}$ are
 easily computable~\footnote{One can also compute all the non-gaussian parameters, $f_{NL}^{\mathrm{local}},\tau_{NL}^{\mathrm{local}},g_{NL}^{\mathrm{local}}$ 
and CMB dipolar asymmetry parameter, $A_{CMB}$ using {\it In-In formalism} in the quntum regime. But the inflationary 
dynamics responsible for the interactions between the
modes occurs at the super-horizon scales within the effective theory setup proposed in this paper. 
Here I use $\delta N$ formalism as-(1) it perfectly holds good at the super-horizon scales
and (2) are also independent of any kind of intrinsic non-Gaussianities generated at the scale of horizon
crossing.}.

 We will be using the following constraints on the amplitude of the power spectrum, $P_s$, spectral tilt, $n_s$, tensor-to-scalar ratio, $r$, sound speed, $c_{s}$, local type
 of non-Gaussianity, $f_{NL}^{\mathrm{local}}$ and $\tau_{NL}^{\mathrm{local}}$ and CMB dipolar asymmetry from Planck data
 throughout the paper \cite{Ade:2013uln,Ade:2013zuv,Ade:2013ydc,Ade:2013nlj}:

\begin{eqnarray}
\label{pow} \displaystyle \ln(10^{10}P_{s})=3.089^{+0.024}_{-0.027}~~(~at~2\sigma~~ CL)\,, \\
\label{ns} \displaystyle  n_{s}=0.9603\pm 0.0073~~(~at~2\sigma~~ CL)\,, \\
\label{rten} \displaystyle r\leq 0.12~~(~at~2\sigma~~ CL)\,, \\
\label{f-nl}
\displaystyle 0.02 \leq c_s \leq 1~~(~at~2\sigma~~ CL)\,, \\
\label{f-nl1}\displaystyle f_{NL}^{\mathrm{local}}=2.7\pm 5.8~~(~at~1\sigma~~ CL)\,,
\\ \label{f-nl2}\displaystyle \tau_{NL}^{\mathrm{local}}\leq2800~~(~at~2\sigma~~ CL)\,, \\
\label{f-nl3}\displaystyle A_{CMB}=0.07\pm 0.02~~(~at~2\sigma~~ CL).
\end{eqnarray}

In this paper I will concentrate our study for Hubble induced inflection point MSSM inflation derived from various higher dimensional Planck scale suppressed 
non-minimal K\"ahler operators
in ${\cal N}=1$ supergravity (SUGRA) which satisfies the observable universe, and it is well motivated for providing an example of visible sector inflation.

In section \ref{setup}, I will briefly review the setup with one heavy and one light superfield which are coupled via non-minimal interactions through
 K\"ahler potential. In section \ref{N-M} I discuss very briefly the role of various types of Planck suppressed non-minimal K\"ahler
 corrections to model a Hubble induced MSSM inflation for any D-flat directions. Hence in section \ref{BNV} I present a quantitaive
 analysis to compute the expression for the local types of non-Gaussianin parameters and CMB dipolar asymmetry
parameter which characterize the 
bispectrum and trispectrum using the $\delta N$ formalism. For the numerical estimations I analyze the results in the context of 
two D-flat direction, $\widetilde L\widetilde L\widetilde e$ and $\widetilde u\widetilde d\widetilde d$ within
 the framework of MSSM inflation \cite{Allahverdi:2006iq,Allahverdi:2006cx}.


\section{Planck suppressed non-minimal K\"ahler operators within ${\cal N}=1$ SUGRA}\label{setup}

\subsection{ The Superpotential}
In this section I concentrate on two sectors; heavy hidden sector denoted by the superfield $S$, 
and the light visible sector denoted by $\Phi$ where they interact only via gravitation. 
Specifically the inflaton superfield $\Phi$ is made up of $D$-flat 
direction within MSSM and they are usually lifted by the $F$-term \cite{Enqvist:2003gh}
of the non-renormalizable operators as appearing in the superpotential.
In the present setup for the simplest situation I start with the following simplified expression for the superpotential made up of 
the superfields $S$ and $\Phi$ as given by:
\begin{eqnarray}
 W(\Phi,S)&=&W(\Phi)+W(S)\,=\frac{\lambda \Phi^n}{n M_{p}^{n-3}}+\frac{M_s}{2}S^2\,,
\end{eqnarray}
where for MSSM D-flat directions, $n\geq 3$ (In the present context $n$ characterizes the dimension of the non-renormalizable operator) and the coupling, $\lambda\sim {\cal O}(1)$.
The scale $M_s$ characterizes the scale of heavy physics which belongs to the hidden sector of the effective theory.
 Furthermore, I will assume that the VEV, $\langle s \rangle=M_{s}\leq M_{p}$ and $\langle \phi\rangle=\phi_{0}  \leq M_{p}$, 
where both $s$ and $\phi$ are fields corresponding
to the super field $S$ and $\Phi$. We also concentrate on two MSSM flat directions, $\widetilde L\widetilde L\widetilde e$ and $\widetilde u\widetilde d\widetilde d$, 
which can drive inflation with $n=6$ via $R$-parity invariant $(\widetilde L\widetilde L\widetilde e)(\widetilde L\widetilde L\widetilde e)/M^{3}_{p}$
 and $(\widetilde u\widetilde d\widetilde d)(\widetilde u\widetilde d\widetilde d)/M^{3}_{p}$ operators in the visible sector, which are lifted by 
themselves~\cite{Dine:1995kz}, where $\widetilde u,~\widetilde d$ denote the right handed squarks, and $\widetilde L$ denotes that left handed sleptons and 
$\widetilde e$ denotes the right handed slepton. 
 
\begin{figure}[t]
{\centerline{\includegraphics[width=15.5cm, height=12.2cm] {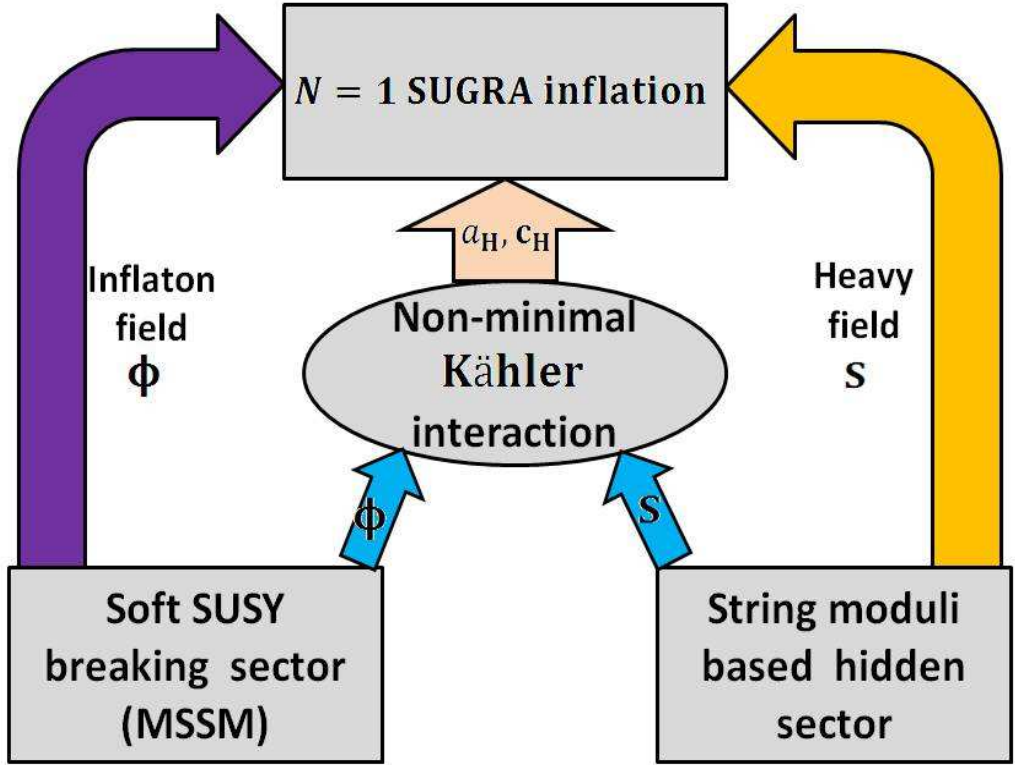}}}
\caption{Schematic representation of effective field theory setup within ${\cal N}=1$ SUGRA setup in presence of non-minimal K\"ahler interaction.
In the final step ${\cal N}=1$ SUGRA inflation is governed by the light inflaton field $\phi$ which belongs to the soft SUSY breaking MSSM sector.
Additionally, the sinusoidally time dependent
dynamical heavy field $s$ always triggers the dynamics of the inflationary framework as the VEV, $\langle s \rangle=M_{s}(\neq 0)<M_{p}$ via the Hubble 
induced correction $\langle V(s) \rangle=M^{4}_{s}$ and non-minimal interactions $a,b,c,d$ as appearing in the $a_{H}$ (A-term) and $c_{H}$ (mass term) in the
SUGRA induced MSSM inflation.} \label{fig3c}
\end{figure}

\subsection{ The K\"ahler potential}
In this paper I consider the following simplest choice of the holomorphic K\"ahler potential
which produces minimal kinetic term,  and the K\"ahler correction of the form:
\begin{eqnarray}\label{klj}
K(s,\phi,s^{\dagger},\phi^{\dagger})=\underbrace{s^{\dagger}s+\phi^{\dagger}\phi}_{minimal~part}+\underbrace{\delta K}_{non-miniml~part}\,,
\end{eqnarray}
where $\delta K$ represent
the higher order non-minimal K\"ahler corrections which are extremely hard to compute from the original string theory background. 
in a more generalized prescription such corrections allow the mixing between the hidden sector heavy fields and the soft SUSY breaking visible sector MSSM fields.  
Using Eq~(\ref{klj}) the most
general ${\cal N} = 1$ SUGRA kinetic term for $(s,\phi)$ field can be written in presence of the non-minimal K\"ahler corrections
through the K\"ahler metric as:
\be\begin{array}{llll}\label{kinf}
    \displaystyle {\cal L}_{Kin}=\left(1+\frac{\partial \delta K}{\partial\phi\partial\phi^{\dagger}}\right)(\partial_{\mu}\phi^{\dagger})(\partial^{\mu}\phi)+
                                  \left(1+\frac{\partial \delta K}{\partial s\partial s^{\dagger}}\right)(\partial_{\mu}s^{\dagger})(\partial^{\mu}s)\\
\displaystyle ~~~~~~~~~~~~~~~~~~~~~~~~~~+\left(\frac{\partial \delta K}{\partial\phi\partial s^{\dagger}}\right)(\partial_{\mu}s^{\dagger})(\partial^{\mu}\phi)+
\left(\frac{\partial \delta K}{\partial s\partial\phi^{\dagger}}\right)(\partial_{\mu}\phi^{\dagger})(\partial^{\mu}s).
   \end{array}
\ee
In this paper I consider the following gauge invariant non-minimal Planck scale suppressed K\"ahler operators within ${\cal N}=1$ SUGRA~\cite{Dine:1995kz,McDonald:1999nc,Kasuya:2006wf}:
\begin{eqnarray}~\label{corrt-1}
 \delta K^{(1)} &=& \frac{a}{M_{p}^2}\phi^\dag \phi s^\dag s + h.c.+\cdots\,, \\
 \label{corrt-2}
 \delta K^{(2)}&=&\frac{b}{2M_{p}}s^\dag\phi \phi + h.c.+\cdots\,, \\
 \label{corrt-3}
 \delta K^{(3)}&=&\frac{c}{4M_{p}^2}s^\dag s^\dag\phi \phi + h.c.+\cdots\,, \\
 \label{corrt-4}
 \delta K^{(4)}&=&\frac{d}{M_{p}}s\phi^\dag \phi + h.c. +\cdots\,,
\end{eqnarray}
where $a,~b,~c,~d$ are dimensionless non-minimal coupling parameters. The $\cdots$ contain higher order non minimal terms which has been ignored in this paper. 
In Fig~(\ref{fig3c}) let me have shown the 
schematic picture of the total effective field theory setup within
 ${\cal N}=1$ SUGRA in presence of non-minimal K\"ahler interaction.

\section{Modeling MSSM inflation from light \& heavy sector}\label{N-M}

In this section let me consider a situation where inflation occurs via the slow roll of $\phi$ field within an MSSM vacuum with a {\it gauge 
enhanced symmetry}, where the entire electroweak symmetry is completely restored.
Let me imagine a physical situation where the heavy field is coherently oscillating around a VEV, $\langle s\rangle \sim M_s$, during the 
initial phase of inflation,
\begin{equation}\label{heavyf}
 s(t)=M_{s}+M_{s}\sin(M_st)\,
 \end{equation}
which arises quite naturally from the hidden sector string moduli field and is coherently oscillating before
 being damped away by the phase of inflation.
The contribution to the potential due to the time dependent oscillating heavy field, with an effective mass $M_s \gg H_{inf}$,
 is averaged over a full cycle ($0<t_{osc}<H_{inf}^{-1}$) is given by:
\be\begin{array}{lll}\label{avg}
    \displaystyle \langle V(s) \rangle \approx M_s^{2} \langle s^{2}(t) \rangle \sim H^{2} M^{2}_{p}.
   \end{array}
\ee
 
Let me now concentrate on the Hubble induced potential when $V(s)=3H^2M^{2}_{p}\sim M^{2}_{s}|s|^2$, in which case the 
contributions from the Hubble-induced terms are important compared to the soft SUSY breaking mass, $m_\phi$, and the $A$ term for 
all the cases mentioned in Eq~(\ref{corrt-1}-\ref{corrt-4}).
Consequently the potential is dominated by a single scale, i.e. $H\sim H_{inf}$, \cite{Choudhury:2014sxa}
\begin{equation}\label{h1a}
 V(\phi)=V(s)+c_{H}H^{2}|\phi|^{2}-
 \frac{a_{H}H\phi^n}{nM_{p}^{n-3}}+\frac{\lvert\phi\rvert^{2(n-1)}}{M_{p}^{2(n-3)}},
\end{equation}
where I have taken $\lambda=1$ and, the Hubble-induced mass parameter $c_{H}$
 and A-term $a_{H}$~\footnote{In the present setup the Hubble induced mass term $c_{H}$ and the A-term $a_{H}$ can
 be expressed in terms of the non-minimal parameters $a,b,c,d$. For more details see Ref~\cite{Dine:1995kz,McDonald:1999nc,Kasuya:2006wf} on these aspects.} for $s<<M_{p}$ can be expressed
 in terms of the non-minimal couplings appearing in Eq~(\ref{corrt-1}).
Fortunately for this class of potential given by Eq~(\ref{h1a}), inflection point inflation can be 
characterized by a fine-tuning parameter, $\delta$, defined as~\cite{Allahverdi:2006we}:
\begin{equation}
\label{newbeta}
\frac{a_H^2}{8(n-1)c_H} = 1-\left(\frac{n-2}{2}\right)^2\delta^2\,.
\end{equation}
Here the tuning parameter, $\vert\delta\vert$ is small enough~\footnote{In the present context I consider that the tuning is of the order of $\delta \sim 10^{-4}$},
for which a point of inflection at the position of the VEV of the inflaton $\langle \phi\rangle=\phi_0$ exists,
so that the flatness condition $V^{\prime\prime}\left(\phi_0\right) =0$ holds good in the present context, with
\begin{equation}
\label{phi0}
\phi_0 = \left(\sqrt{\frac{c_H}{(n-1)}} H M_{p}^{n-3}\right)^{{1}/{n-2}}\, +{\cal O}(\delta^2).
\end{equation}

For $\delta <1$, one can Taylor-expand the inflaton potential around an inflection 
point, $\phi=\phi_{0}$, as~\cite{Allahverdi:2008bt,Enqvist:2010vd,Mazumdar:2011ih,Hotchkiss:2011am,Wang:2013hva,Choudhury:2013jya}:
\be\label{rt1a}
V(\phi)=\alpha+\beta(\phi-\phi_{0})+\gamma(\phi-\phi_{0})^{3}+\kappa(\phi-\phi_{0})^{4}+\cdots\,,
\ee 
where $\alpha$ denotes the height of the potential, and the coefficients $\beta,~\gamma,~\kappa$ determine the shape of the 
potential in terms of the model parameters~\footnote{The analytical expressions for the co-efficients appearing in the {\it inflection point} potential, 
$\alpha,\beta,\gamma$ and $\kappa$, can be expressed in terms of 
the mass parameter $c_{H}$, Hubble scale $H$ and, the VEV of the inflaton $\phi_{0}$ and tuning parameter $\delta$ are explicitly mentioned in the appendix.}.
 Note that once the numerical values of $c_H$ and $H$ are specified, all the terms in the potential are determined. 

For $c_{s}\neq 1$, the upper bound on the numerical value of the Hubble 
 parameter ($H$), the scale of inflation ($\sqrt[4]{V_{*}}$) and the scale of the heavy string moduli field ($M_{s}$) are expressed as:
\begin{equation}\label{hubble}
     H\leq 9.241\times 10^{13}\times\sqrt{\frac{r_{*}}{0.12}}~c^{\frac{\epsilon_{V}}{\epsilon_{V}-1}}_{s}~{\rm GeV}\,,
   \end{equation}
   
\begin{equation}\label{scale}
     \sqrt[4]{V_{*}}\leq 1.96\times 10^{16}\times\sqrt[4]{\frac{r_{*}}{0.12}}~c^{\frac{\epsilon_{V}}{2(\epsilon_{V}-1)}}_{s}~{\rm GeV}\,,
   \end{equation}

\begin{equation}\label{hscalecon1}
     \displaystyle M_{s}\leq 1.77\times 10^{16}\times\sqrt[4]{\frac{r_{*}}{0.12}}~c^{\frac{\epsilon_{V}}{2(\epsilon_{V}-1)}}_{s}~{\rm GeV}\,.
    \end{equation}
where $r_{*}$ is the tensor-to-scalar ratio at the pivot scale of momentum $k_{*}\sim 0.002 Mpc^{-1}$.

\section{ $\delta N$ formalism in presence of non-minimal K\"ahler operators for $c_s\neq 1$}\label{BNV}

In this section I have used the $\delta N$ formalism \cite{Starobinsky:1982,Salopek:1990,Sasaki:1995aw,Wands:2000dp,Lyth:2004gb,Lyth:2005fi,Mazumdar:2012jj,Sugiyama:2012tj}
to compute the local type of non-Gaussianity, $f_{NL}^{\rm local}$ from the
 prescribed setup for $c_s\neq 1$. 
In the non-attractor regime, the $\delta N$
formalism shows various non trivial features which has to be taken into account during explicit calculations. Once the solution
reaches the attractor behaviour, the dominant contribution comes from only on the
perturbations of the scalar-field
trajectories with respect to the inflaton field value at the initial
hypersurface, $\phi$, as the velocity, $\dot\phi$, is uniquely
determined by $\phi$. However, in the non-attractor regime of solution, 
both the information from the field value $\phi$ and also $\dot \phi$ are required to determine the 
trajectory \cite{Namjoo:2012aa}.

\begin{figure}[t]
{\centerline{\includegraphics[width=15.5cm, height=12.2cm] {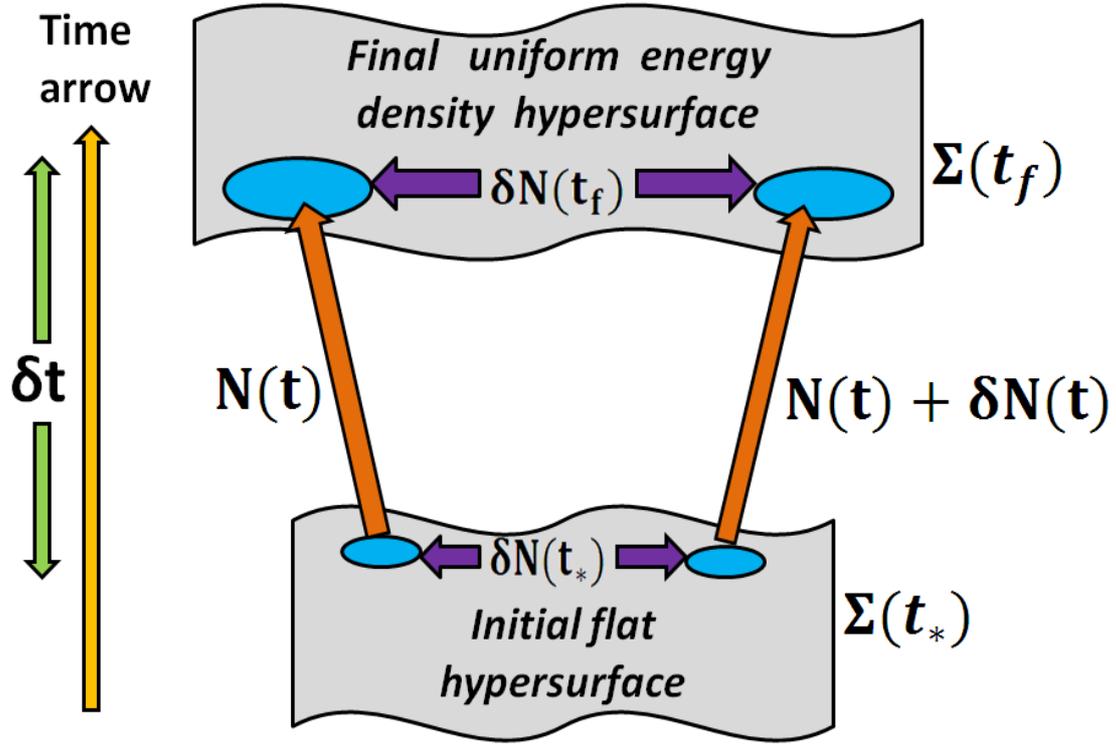}}}
\caption{Diagrammatic representation of $\delta N$ formalism. In this schematic picture $\Sigma(t_{i})$ and $\Sigma(t_{f})$ represent the initial and final hypersurface
where time arrow flows from $t_{i}\rightarrow t_{f}$.} \label{fig3ccz}
\end{figure}

In order to compute the scalar-field trajectories explicitly, here I feel the need to solve the equation
of motion of the scalar field, which is in general a second-order differential equation in a prescribed background. 
This can be solved by providing two initial conditions on $\phi$ and $\dot\phi$ on the
initial hypersurface. I have, therefore, integrated the equation of motion to
the final time, $t=t_*$. Here I have solved the equation of motion 
perturbatively by expanding it around a particular trajectory given by
$\phi\propto e^{\vartheta Ht}$, where $\vartheta$ is time dependent function in the generalized physical prescription. So I have used these background solutions for
the field trajectories to compute the perturbations of the number of
$e$-folds with respect to the initial field value and its time derivative.
During the computation of the trajectories let me assume here that the universe has already arrived 
at the adiabatic limit via attractor phase by this epoch,
or equivalently it can be stated that a typical phase transition phenomenona appears to an attractor phase at the time $t=t_*$. 
More specifically, in the present context, I have assumed that the evolution of the universe is unique
after the value of the scalar field arrived at $\phi=\phi_*$ where it is mimicking the role of standard clock, 
irrespective of the value of its velocity $\dot\phi_*$. 
Let me mention that only in this case $\delta N$ is equal to
the final value of the comoving curvature perturbation $\zeta$ which
is conserved at $t\geq t_*$. In Fig~(\ref{fig3ccz}) I have shown the schematic picture of the $\delta N$ formalism. 
\subsection{General conventions}

In the present context, further I have neglected the canonical kinetic term
during the non-attractor phase for simplicity. The background equation
of motion for the four physical situations are given by
\be\label{svb}
 0=\left\{
	\begin{array}{ll}
                     \small\small \left(\ddot{\phi}+3H\dot{\phi}\right)\left[1+\frac{aM^{2}_{s}}{2M^{2}_{p}}\left(1+\sin(M_{s}t)\right)^{2}\right]
+\frac{2aM^{3}_{s}}{M^{2}_{p}}\left[\left(2\dot{\phi}
+3H\phi\right)\cos(M_{s}t)\right.\\ \left.~~~~~~~~~~~~~~~~~~~~~~~~~~~~~~~~~~~~~~~~~-\phi M_{s}\sin(M_{s}t)\right]\left(1+\sin(M_{s}t)\right)
+V^{'}(\phi) & \mbox{ for $\underline{\bf Case ~I}$}  \\ 
   \ddot{\phi}+3H\dot{\phi}+\frac{bM^{2}_{s}}{M_{p}}\left[\left(\dot{\phi}+3H\phi\right)\cos(M_{s}t)-M_{s}\phi\sin(M_{s}t)\right]+V^{'}(\phi)& \mbox{ for $\underline{\bf Case ~II}$}  \\ 
    \ddot{\phi}+3H\dot{\phi}+\frac{cM^{3}_{s}}{2M^{2}_{p}}\left[\left(\dot{\phi}+3H\phi\right)\cos(M_{s}t)\left(1+\sin(M_{s}t)\right)+M_{s}\phi\left(\cos^{2}(M_{s}t)
\right.\right.\\ \left.\left.~~~~~~~~~~~~~~~~~~~~~~~~~~~~~~~~~~~~~~~~~~~~-\sin(M_{s}t)\left(1+\sin(M_{s}t)\right)\right)\right]+V^{'}(\phi) & \mbox{ for $\underline{\bf Case ~III}$}  \\ 
    \left(\ddot{\phi}+3H\dot{\phi}\right)\left[1+\frac{2dM_{s}}{M_{p}}\left(1+\sin(M_{s}t)\right)\right]
+\frac{2dM^{2}_{s}}{M_{p}}\left[\left(2\dot{\phi}
+3H\phi\right)\cos(M_{s}t)\right.\\ \left.~~~~~~~~~~~~~~~~~~~~~~~~~~~~~~~~~~~~~~~~~~~~~~~~~~~~~~~~~-\phi M_{s}\sin(M_{s}t)\right]
+V^{'}(\phi)      & \mbox{ for $\underline{\bf Case ~IV}$}.
          \end{array}
\right.
\ee
From the Eq~(\ref{svb}), it is obvious that the determination of a general analytical solution is
too much complicated. To simplify the task here I consider a particular solution,
\be\phi=\phi_{L}\propto e^{\vartheta Ht} ~~~(~{\rm i.e.}~~~ \phi=\phi_{L}(N)=\phi_*e^{-\vartheta N}),\ee
and further my prime objective is to obtain a more generalized solution for the background up to the
second order in perturbations around this particular solution. 
Here I also assume that the non-attractor phase ends when the inflaton field value is achieved at $\phi=\phi_*$.
Let me define a perturbative parameter,
$$\Delta\equiv\phi-\phi_{0}-\phi_{L}\,=\Delta_{1}+\Delta_{2}+\cdots , $$ which represents the difference
between the true background solution and the reference solution to solve the background Eq~(\ref{svb}) perturbatively. Here 
$\Delta_{1}$ and $\Delta_{2}$ are the general linearized and second order perturbative solution of the background field equations.
The $\cdots$ contribution comes from the higher order perturbation scenario which I will neglect for further computation.

\subsection{Linearized perturbation}

Let me consider the contribution from the linear perturbation, $\Delta_{1}$. Consequently in the leading order the background linearized perturbative equation of
motion takes the following form:
\be\label{svblin}
 0\approx\left\{
	\begin{array}{ll}
                     \small\small \left(\ddot{\Delta_{1}}+3H\dot{\Delta_{1}}+\vartheta H^{2}(3+\vartheta)\phi_{L}\right)
\left[1+\frac{aM^{2}_{s}}{4M^{2}_{p}}\right]
-\frac{aM^{3}_{s}}{M^{2}_{p}}(\Delta_{1}+\phi_{L}) M_{s}
+\beta & \mbox{ for $\underline{\bf Case ~I}$}  \\ 
   \ddot{\Delta_{1}}+3H\dot{\Delta_{1}}+\vartheta H^{2}(3+\vartheta)\phi_{L}+\beta & \mbox{ for $\underline{\bf Case ~II}$}  \\ 
    \ddot{\Delta_{1}}+3H\dot{\Delta_{1}}+\vartheta H^{2}(3+\vartheta)\phi_{L}+\beta & \mbox{ for $\underline{\bf Case ~III}$}  \\ 
    \left(\ddot{\Delta_{1}}+3H\dot{\Delta_{1}}+\vartheta H^{2}(3+\vartheta)\phi_{L}\right)\left[1+\frac{2dM_{s}}{M_{p}}\right]
+\beta   & \mbox{ for $\underline{\bf Case ~IV}$}.
          \end{array}
\right.
\ee
where I have neglect the higher powers of $\Delta_{1}$ in the linearized approximation.
The general solution is given by
\be\label{svblin2}
 \Delta_{1}\approx\left\{
	\begin{array}{ll}
                     \small\small {\bf C}_{1}e^{\frac{1}{2}
\left(-3H-\sqrt{\frac{4aM^{4}_{s}}{M^{2}_{p}\left(1+\frac{aM^{2}_{s}}{4M^{2}_{p}}\right)}+9H^{2}}\right)t}
+{\bf C}_{2}e^{\frac{1}{2}
\left(-3H+\sqrt{\frac{4aM^{4}_{s}}{M^{2}_{p}\left(1+\frac{aM^{2}_{s}}{4M^{2}_{p}}\right)}+9H^{2}}\right)t}\\
~~~~~~~~~~~~~~~~~~~~~~~~~~~~~~~~~~~~~~~~~~~~~~~~~~~~~~+\phi_{*}e^{\vartheta Ht}
-\frac{\beta M^{2}_{p}}{aM^{4}_{s}} & \mbox{ for $\underline{\bf Case ~I}$}  \\ 
  {\bf C}_{3}-\frac{{\bf C}_{4}}{3H}e^{-3Ht}-\frac{\beta t}{3H}-\phi_{*}e^{\vartheta Ht} & \mbox{ for $\underline{\bf Case ~II}$}  \\ 
    {\bf C}_{5}-\frac{{\bf C}_{6}}{3H}e^{-3Ht}-\frac{\beta t}{3H}-\phi_{*}e^{\vartheta Ht} & \mbox{ for $\underline{\bf Case ~III}$}  \\ 
    {\bf C}_{7}-\frac{{\bf C}_{8}}{3H}e^{-3Ht}-\frac{\beta t}{3H\left(1+\frac{2dM_{s}}{M_{p}}\right)}-\phi_{*}e^{\vartheta Ht}  & \mbox{ for $\underline{\bf Case ~IV}$}.
          \end{array}
\right.
\ee
where ${\bf C}_{i}\forall i(=1,2,....,8)$ are dimensionful arbitrary integration constants which can be fixed by imposing the boundary conditions.

\subsection{Second-order perturbation}

Next I have considered the contribution from the second-order perturbation, $\Delta_2$.
 Consequently in the leading order the background Second-order perturbative equation of
motion takes the following form:
\be\label{svbsec2}
 \Pi_{s}\approx\left\{
	\begin{array}{ll}
                     \small\small \left(\ddot{\Delta_{2}}+3H\dot{\Delta_{2}}+\vartheta H^{2}(3+\vartheta)\phi_{L}\right)
\left[1+\frac{aM^{2}_{s}}{4M^{2}_{p}}\right]
-\frac{aM^{3}_{s}}{M^{2}_{p}}(\Delta_{2}+\phi_{L}) M_{s}
+\beta & \mbox{ for $\underline{\bf Case ~I}$}  \\ 
   \ddot{\Delta_{2}}+3H\dot{\Delta_{2}}+\vartheta H^{2}(3+\vartheta)\phi_{L}+\beta & \mbox{ for $\underline{\bf Case ~II}$}  \\ 
    \ddot{\Delta_{2}}+3H\dot{\Delta_{2}}+\vartheta H^{2}(3+\vartheta)\phi_{L}+\beta & \mbox{ for $\underline{\bf Case ~III}$}  \\ 
    \left(\ddot{\Delta_{2}}+3H\dot{\Delta_{2}}+\vartheta H^{2}(3+\vartheta)\phi_{L}\right)\left[1+\frac{2dM_{s}}{M_{p}}\right]
+\beta   & \mbox{ for $\underline{\bf Case ~IV}$}.
          \end{array}
\right.
\ee
where the source term, $\Pi_{s}$, for the sub-Planckian Hubble induced inflection point inflation within ${\cal N}=1$ SUGRA is given by
\be
\Pi_{s}=3\gamma(\Delta_{1}+\phi_{L})^{2}\,.
\ee
Now to solve Eq~(\ref{svbsec2}) in presence of non-linear source term, let me assume that the contribution from $\phi_{L}$ is sub-dominant.
Consequently the general solution in presence of second-order perturbation is given by:
\be\label{svbper4}
 \Delta_{2}\approx\left\{
	\begin{array}{ll}
                     \small\small {\bf G}_{1}e^{\frac{1}{2}
\left(-3H-\sqrt{\frac{4aM^{4}_{s}}{M^{2}_{p}\left(1+\frac{aM^{2}_{s}}{4M^{2}_{p}}\right)}+9H^{2}}\right)t}
+{\bf G}_{2}e^{\frac{1}{2}
\left(-3H+\sqrt{\frac{4aM^{4}_{s}}{M^{2}_{p}\left(1+\frac{aM^{2}_{s}}{4M^{2}_{p}}\right)}+9H^{2}}\right)t}+\Sigma_{s}(t) & \mbox{ for $\underline{\bf Case ~I}$}  \\ 
  {\bf G}_{5}-\frac{12{\bf G}_{6}}{H}e^{-3Ht}+\Xi_{s}(t) & \mbox{ for $\underline{\bf Case ~II}$}  \\ 
     {\bf G}_{5}-\frac{12{\bf G}_{6}}{H}e^{-3Ht}+\Psi_{s}(t) & \mbox{ for $\underline{\bf Case ~III}$}  \\ 
    {\bf G}_{7}-\frac{12{\bf G}_{8}}{H}e^{-3Ht}+\Theta_{s}(t) & \mbox{ for $\underline{\bf Case ~IV}$}.
          \end{array}
\right.
\ee
where the time dependent functions $\Sigma_{s}(t),\Xi_{s}(t),\Psi_{s}(t)$ and $\Theta_{s}(t)$ are explicitly mentioned in the Appendix \ref{time dep}.
Here ${\bf G}_{i}\forall i(=1,2,....,8)$ are dimensionful arbitrary integration constants which can be fixed by imposing the boundary conditions.
\subsection{$\delta N$ at the final hypersurface}

In the present context my prime objective is to compute the perturbations of the number of
$e$-folds, $\delta N$.
The truncated background solution of $\phi$ up to the
second-order perturbations around the reference trajectory, 
$\phi_L\propto e^{-\vartheta N}$ in terms of $N$ is given by,
\be
\phi =\phi_{0}+\left\{
	\begin{array}{ll}
              \frac{\phi_*}{1+\widehat{{\bf C}}_{1}+\widehat{{\bf C}}_{2}-\frac{\beta M^{2}_{p}}{a \phi_{*}M^{4}_{s}}
+\widehat{{\bf G}}_{1}+\widehat{{\bf G}}_{2}+\widehat{\Sigma}_{s}(0)}\,
\left(e^{-\vartheta N} +\widehat{\Delta}_{1}(N)
 +\widehat{\Delta}_{2}(N)\right)\,& \mbox{ for $\underline{\bf Case ~I}$}  \\ 
   \frac{\phi_*}{1+\widehat{{\bf C}}_{3}-\frac{\widehat{{\bf C}}_{4}}{3H}+\widehat{{\bf G}}_{3}-\frac{12\widehat{{\bf G}}_{4}}{H}
+\widehat{\Xi}_{s}(0)}\,
\left(e^{-\vartheta N} +\widehat{\Delta}_{1}(N)
 +\widehat{\Delta}_{2}(N)\right)\, & \mbox{ for $\underline{\bf Case ~II}$}  \\ 
    \frac{\phi_*}{1+\widehat{{\bf C}}_{5}-\frac{\widehat{{\bf C}}_{6}}{3H}+\widehat{{\bf G}}_{5}-\frac{12\widehat{{\bf G}}_{6}}{H}
+\widehat{\Psi}_{s}(0)}\,
\left(e^{-\vartheta N} +\widehat{\Delta}_{1}(N)
 +\widehat{\Delta}_{2}(N)\right)\, & \mbox{ for $\underline{\bf Case ~III}$}  \\ 
    \frac{\phi_*}{1+\widehat{{\bf C}}_{7}-\frac{\widehat{{\bf C}}_{8}}{3H}+\widehat{{\bf G}}_{7}-\frac{12\widehat{{\bf G}}_{8}}{H}
+\widehat{\Theta}_{s}(0)}\,
\left(e^{-\vartheta N} +\widehat{\Delta}_{1}(N)
 +\widehat{\Delta}_{2}(N)\right)\,   & \mbox{ for $\underline{\bf Case ~IV}$}.
          \end{array}
\right.,
\label{phi-lambda2}
\ee
where the symbol $~\widehat{}~$ is introduced to rescale the integration constants as well as the perturbative solutions by $\phi_{*}$.
Here I have neglected the contribution from $e^{-\vartheta N}$ in $\widehat{\Delta}_{2}(N)$ to avoid over counting in the Eq~(\ref{phi-lambda2}).
It is important to note that in the present context of all 
 these sets of scaled
integration constants parameterizes
different trajectories, and I have set $\phi(0,\widehat{{\bf W}}_{k})=\phi_*$
for any value of $\widehat{{\bf W}}_{k}\forall k=([1,2],[3,4],[5,6],[7,8])$ in accordance with the assumption that 
the end of the non-attractor phase is determined only by the value of the
scalar field, $\phi=\phi_*$. Here $\widehat{{\bf W}}_{k}=\widehat{{\bf C}}_{k},\widehat{{\bf G}}_{k}$
represent collection of all integration constants.

Further inverting Eq~(\ref{phi-lambda2}) for a fixed set of $\widehat{{\bf W}}_{k}$, I have obtained
$N$ as a implicit function of $\phi$ and $\widehat{{\bf W}}_{k}$. Then the $\delta N$ formula can be
obtained by \cite{Chen:2013eea}:
\begin{eqnarray}
\delta N=N(\phi+\delta\phi,\widehat{{\bf W}}_{k})-N(\phi,0)
=\sum_{k}\sum_{n,m}\frac{1}{n!m!}
\partial^{n}_{\phi^{n}}\partial^{m}_{\widehat{{\bf W}}^{m}_{k}}N(\phi,0)
\delta\phi^n\widehat{{\bf W}}^{m}_{k}\,.
\end{eqnarray}
Here I have introduced the shift in the inflaton field $\phi\to\phi+\delta\phi$ and the number of e-folds $N\to N+\delta N$
on both sides of Eq~(\ref{phi-lambda2}) to compute $\delta N$ iteratively.
In the present context I have obtained, perturbative
solutions of the scalar-field trajectories around the particular
reference solution, $\phi_L=\phi_*e^{\vartheta Ht}$, which are
valid only when the perturbed trajectories are not far away from the
reference solution. Additionally, since I have neglected the sub-dominant solution,
$\Delta_1\propto e^{\vartheta Ht}$, my approximation holds good only at sufficiently 
late times. These imply that here I should choose the initial 
time as close as possible to the final time for which $N\lesssim 1$. 
Then the simplest choice is
to take the initial time to be infinitesimally close to $t=t_*$.

Now perturbing the number of e-folds $N$ up to the second order at the epoch $t=t_*$, I get \cite{Chen:2013eea}:
\begin{eqnarray}\label{efolN}
\zeta=\delta N =  N_{,\phi}\delta\phi
+  \frac12  N_{,\phi\phi}\delta\phi^2
+ \frac16  N_{,\phi\phi\phi}\delta\phi^3 +\cdots .~~~~
\end{eqnarray}
where I have used the $\widehat{{\bf W}}$-independence of $N$ at $N=0$ for which, $N_{,\widehat{{\bf W}}_{k}}=0=N_{,\widehat{{\bf W}}_{p}\widehat{{\bf W}}_{q}}$.
Here $\cdots$ corresponds to the higher order contributions, which are negligibly small compared to the leading order contributions.
By taking the derivatives of both sides of Eq.~(\ref{phi-lambda2})
and setting $N=0=\widehat{{\bf W}}_{k}(=\widehat{{\bf C}}_{k},\widehat{{\bf G}}_{k})\forall k$ at the end,
my next task is to identify $\delta\phi_*$ and $\widehat{{\bf W}}_{k}(=\widehat{{\bf C}}_{k},\widehat{{\bf G}}_{k})$
which are actually generated from quantum fluctuations on flat slicing, $\delta\phi$.
To serve this purpose let me consider the evolution of $\delta\phi$ on super-horizon scales.
The shift in the inflaton field can be expressed here as:
\be\begin{array}{llll}\label{eqrt1}
    \displaystyle \delta\phi(N)=\delta\phi_{1}(N)+\delta\phi_{2}(N)=\phi_{*}\left(\widehat{\Delta}_{1}(N)+\widehat{\Delta}_{2}(N)\right)
   \end{array}\ee
where the subscript ``1'' and ``2'' represent the solution at the linear and the second order respectively.
It is important to note that both the solutions include the features of growing and decaying mode.
Now imposing the boundary condition from the end of the non-attractor
phase, where $N=0$, I get:
\be
\delta\phi_1(0)=\delta\phi_{1*}=\phi_{*}\widehat{\Delta}_{1}(0)\,=\left\{
	\begin{array}{ll}
                     \small\small \phi_{*}\left(\widehat{{\bf C}}_{1}
+\widehat{{\bf C}}_{2}
-\frac{\beta M^{2}_{p}}{a\phi_{*}M^{4}_{s}}\right) & \mbox{ for $\underline{\bf Case ~I}$}  \\ 
  \phi_{*}\left(\widehat{{\bf C}}_{3}-\frac{\widehat{{\bf C}}_{4}}{3H}\right) & \mbox{ for $\underline{\bf Case ~II}$}  \\ 
    \phi_{*}\left(\widehat{{\bf C}}_{5}-\frac{\widehat{{\bf C}}_{6}}{3H}\right) & \mbox{ for $\underline{\bf Case ~III}$}  \\ 
    \phi_{*}\left(\widehat{{\bf C}}_{7}-\frac{\widehat{{\bf C}}_{8}}{3H}\right)  & \mbox{ for $\underline{\bf Case ~IV}$}.
          \end{array}
\right.
\label{dphistar}
\ee
\be
\delta\phi_2(0)=\delta\phi_{2*}=\phi_{*}\widehat{\Delta}_{2}(0)\,=\left\{
	\begin{array}{ll}
                     \small\small \phi_{*}\left(\widehat{{\bf G}}_{1}
+\widehat{{\bf G}}_{2}
+\widehat{\Sigma}_{s}(0)\right) & \mbox{ for $\underline{\bf Case ~I}$}  \\ 
  \phi_{*}\left(\widehat{{\bf G}}_{3}-\frac{12\widehat{{\bf G}}_{4}}{H}+\widehat{\Xi}_{s}(0)\right) & \mbox{ for $\underline{\bf Case ~II}$}  \\ 
    \phi_{*}\left(\widehat{{\bf G}}_{5}-\frac{12\widehat{{\bf G}}_{6}}{H}+\widehat{\Psi}_{s}(0)\right) & \mbox{ for $\underline{\bf Case ~III}$}  \\ 
    \phi_{*}\left(\widehat{{\bf G}}_{7}-\frac{12\widehat{{\bf G}}_{8}}{H}+\widehat{\Theta}_{s}(0)\right)  & \mbox{ for $\underline{\bf Case ~IV}$}.
          \end{array}
\right.
\label{dphistar2}
\ee
from which I have obtained:
\begin{eqnarray}
\delta\phi_*=\delta\phi(0)=\delta\phi_{1*}+\delta\phi_{2*}=\phi_{*}\left(\widehat{\Delta}_{1}(0)+\widehat{\Delta}_{2}(0)\right)\,.
\label{fluctuations}
\end{eqnarray}

Further neglecting the mixing between the solutions corresponding to the
 linearized and second order perturbation,
the analytical expression for $\delta N$ can be expressed as: 
\be\begin{array}{lll}\label{delbn}\displaystyle
\displaystyle \zeta=\delta N = -\dfrac{(\delta\phi_{1*}+\delta\phi_{2*})}{\vartheta \phi_*}+\frac{(\delta\phi^{2}_{1*}+\delta\phi^{2}_{2*})}{2\vartheta\phi^{2}_{*}}+\cdots
\end{array}\ee

\subsection{Computation of local type of non-Gaussianity and CMB dipolar asymmetry}

The local type of non-Gausiianity is originally implemented as a position space expansion of non-Gaussian fluctuations around Gaussian pertabations \cite{Wang:2013zva}:
\be\begin{array}{llll}\label{nong1}
    \displaystyle \zeta({\bf x})=\zeta_{g}({\bf x})+\frac{3}{5}f_{NL}\zeta^{2}_{g}({\bf x})+\frac{9}{25}g_{NL}\zeta^{3}_{g}({\bf x})+\cdots,
   \end{array}
\ee
where $\zeta_{g}({\bf x})$ satisfies the Gaussian statistics. Here $f_{NL}$, $g_{NL}$, $\cdots$ are the non-Gaussian estimators. However, the inflationary perturbation 
itself is implemented in the momentum space and thus the momentum space correlators provide a clear picture of non-Gaussianity compared to the isolated position space.
Using Eq~(\ref{nong1}) in fourier space the three point and the four point correlator can be expressed as \cite{Wang:2013zva}:
\be\begin{array}{llll}\label{nong2}
    \displaystyle \langle \zeta_{\bf k_{1}}\zeta_{\bf k_{2}}\zeta_{\bf k_{3}} \rangle = (2\pi)^{7}\delta^{3}({\bf k_{1}+k_{2}+k_{3}})
\left[\frac{3f_{NL}^{\mathrm{local}}}{10k^{3}_{1}k^{3}_{2}}P_{s}(k_{1})P_{s}(k_{2})+(k_{2}\leftrightarrow k_{3})+(k_{1}\leftrightarrow k_{3})\right]
   \end{array}\ee
  
\be\begin{array}{llll}\label{nong3}
    \displaystyle \langle \zeta_{\bf k_{1}}\zeta_{\bf k_{2}}\zeta_{\bf k_{3}}\zeta_{\bf k_{4}} \rangle = (2\pi)^{9}\delta^{4}({\bf k_{1}+k_{2}+k_{3}+k_{4}})
\left[\frac{27g_{NL}^{\mathrm{local}}}{100}\sum^{3}_{i<j<p =1}\frac{P_{s}(k_{i})P_{s}(k_{j})P_{s}(k_{p})}{(k_{i}k_{j}k_{p})^{3}}\right.\\ \left. 
\displaystyle ~~~~~~~~~~~~~~~~~~~~~~~~~~~~~~~~~~~~~~~~~~~~~~~~~~~~~~~~~~~~~
+\frac{\tau_{NL}^{\mathrm{local}}}{8}\sum^{11}_{j<p,i\neq j,p=1}\frac{P_{s}(k_{ij})P_{s}(k_{j})P_{s}(k_{p})}{(k_{ij}k_{j}k_{p})^{3}}\right]
   \end{array}\ee
My next job is to derive and to estimate the amount of the local type of non-Gaussianity from the prescribed setup. 
Further using the results obtained for $\delta N$ in the earlier section,
 the non-Gaussian parameter corresponding to the local type of non-Gaussianity $f_{NL}^{\rm local}$
~\footnote{In \cite{Chen:2013eea} the authors have shown that for a specific $P(X,\phi)$ theory with $c_{s}\neq 1$ the non-Gaussian parameter,
$f_{NL}^{\rm local}=\frac{5}{4}\left(1+\frac{1}{c^{2}_{s}}\right)$. In this paper I have obtained different result as the non-minimal K\"ahler
interactions within ${\cal N}=1$ SUGRA effective theory setup which is completely different compared to the case studied in \cite{Chen:2013eea}.}, $g_{NL}^{\rm local}$ and $\tau_{NL}^{\rm local}$ can be computed as:
\be\begin{array}{lll}\label{d1}\displaystyle
\displaystyle f_{NL}^{\mathrm{local}} = \frac{5}{6}
\frac{N_{,\phi\phi}}{N^{2}_{,\phi}}+\cdots
=\frac{5\vartheta}{6}+\cdots\,
\end{array}\ee

\be\begin{array}{lll}\label{d11}\displaystyle
\displaystyle \tau_{NL}^{\mathrm{local}} =
\frac{N^{2}_{,\phi\phi}}{N^{4}_{,\phi}}+\cdots
= \vartheta^{2}+\cdots\,
\end{array}\ee

\be\begin{array}{lll}\label{d111}\displaystyle
\displaystyle g_{NL}^{\mathrm{local}} =
\frac{25}{54}\frac{N_{,\phi\phi\phi}}{N^{3}_{,\phi}}+\cdots
= \frac{25\vartheta^{2}}{108}+\cdots\,
\end{array}\ee
where the parameter $\vartheta$, appearing in all the physical situations, 
can be expressed in terms of the sound speed ($c_{s}$), potential dependent slow roll parameter $(\epsilon_{V},\eta_{V})$ and the model parameters $(\alpha,\beta,\gamma)$ as:
\be\label{cx1}
\vartheta\approx\left[\eta_{V}\left(1+\frac{1}{c^{2}_{s}}\right)^{2}+\epsilon_{V}\left(1-\frac{1}{c^{4}_{s}}\right)\right]\approx\left[\frac{6\gamma\phi_{*}M^{2}_{p}}{\alpha}\left(1+\frac{1}{c^{2}_{s}}\right)^{2}
+\frac{\beta^{2}M^{2}_{p}}{2\alpha^{2}}\left(1-\frac{1}{c^{4}_{s}}\right)\right].
\ee

where the sound speed $c_{s}$ can be expressed in terms of non-canonical K\"ahler corrections, $a,b,c,d$ and the scale of heavy field, $M_{s}$,
as:
%
\be\label{sigbz}
 c_s\approx \left[\sqrt{\frac{{\bf \Sigma}_{1}(t)-{\bf \Sigma}_{2}(t)-\dot{\widehat{V}}}{{\bf \Sigma}_{1}(t)+{\bf \Sigma}_{3}(t)+\dot{\widehat{V}}}}\right]_{t=t_{*}}
\ee
%
The dot denotes derivative w.r.t. physical time, $t$. Here $\widehat{V}=V(\phi)-V(S)$ and the symbol
 $\Sigma= X,Y,Z,W$, appearing for the four cases in Eq~(\ref{sigbz}) are mentioned in 
the appendix. 

Additionally, it is important to note that
the well-known {\it Suyama-Yamaguchi} consistency relation \cite{Suyama:2007bg,Ichikawa:2008iq} between
 the three and four point non-Gaussian parameters, $f_{NL}^{\mathrm{local}}$, $\tau_{NL}^{\mathrm{local}}$ and $g_{NL}^{\mathrm{local}}$ 
 violates \cite{Smith:2011if,Sugiyama:2011jt,Choudhury:2012kw,Rodriguez:2013cj} in the present context due to the appearance of $\cdots$ terms in Eq~(\ref{d1},\ref{d11},\ref{d111}). As the contributions
 form $\cdots$ terms are positive, the consistency relation is modified as:
\be\label{conq}
g_{NL}^{\mathrm{local}}=\frac{25}{108}\tau_{NL}^{\mathrm{local}}=\frac{9}{25}\left(f_{NL}^{\mathrm{local}}\right)^{2}+\cdots.\ee
However, it is important to note that since $\cdots$ terms are small, the amount of violation is also small.

Further using $\delta N$ formalism, the CMB dipolar asymmetry parameter for single field inflationary framework can be expressed as \cite{Kohri:2013kqa}:

\be\begin{array}{llll}\label{eqaq}
    \displaystyle A_{CMB}=\frac{1}{4}\frac{\Delta P_{s}(k)}{P_{s}(k)}\approx \frac{1}{2}\frac{\Delta (\delta N)}{\delta N}=
\frac{3}{5}f_{NL}^{\mathrm{local}}|N_{,\phi}\Delta\phi|+\frac{27}{50}g_{NL}^{\mathrm{local}}|N_{,\phi}\Delta\phi|^{2}+\cdots
   \end{array}\ee
where $|N_{,\phi}\Delta\phi|<1$ for which the perturbative expansion is valid here. In it the field excursion
 $\Delta\phi=\phi_{cmb}-\phi_{e}\approx \phi_{*}-\phi_{e}$, can be expressed as \cite{Choudhury:2013iaa,Choudhury:2013woa}:
\be\begin{array}{llll}\label{fexc}
    \displaystyle \frac{|\Delta \phi|}{M_{p}}\approx \frac{3}{25\sqrt{c_{s}}}\sqrt{\frac{r_{*}}{0.12}}\left|\left\{\frac{3}{400}
\left(\frac{r_{*}}{0.12}\right)-\frac{\eta_{V}(k_{\star})}{2}-\frac{1}{2}
\,\right\}\right|
   \end{array}\ee

where $\phi_{cmb}\approx \phi_*$ and $\phi_{e}$ be the value of the inflaton field at the horizon crossing and the end of inflation respectively. 
Here $r_{*}$ be the tensor-to-scalar ratio
at the pivot scale of momentum, $k_{*}\sim 0.002~{\rm Mpc}^{-1}$. Hence substituting Eq~(\ref{d1},\ref{d11},\ref{d111}) in Eq~(\ref{eqaq}) I derive
the following simplified expression in terms of the tensor-to-scalar ratio, sound speed and :
\be\begin{array}{llll}\label{eqaq1}
    \displaystyle A_{CMB}=
\frac{1}{2}\frac{|\Delta\phi|}{\phi_*}+\frac{1}{8}\left(\frac{|\Delta\phi|}{\phi_*}\right)^{2}+\cdots
\\
~~~~~~~~\approx \displaystyle \frac{3M_{p}}{50\phi_*\sqrt{c_{s}}}\sqrt{\frac{r_{\star}}{0.12}}\left|\left\{\frac{3}{400}
\left(\frac{r_{\star}}{0.12}\right)-\frac{3\gamma\phi_{*}M^{2}_{p}}{\alpha}-\frac{1}{2}
\,\right\}\right|+\cdots
   \end{array}\ee
where $\frac{|\Delta\phi|}{\phi_*}<1$ in the present sub-Planckian setup.

\subsection{Constraining local type of non-Gaussianity and CMB dipolar asymmetry via multi parameter scanning}

\begin{figure*}[htb]
\centering
\subfigure[$\textcolor{blue}{\bf \underline{Case ~I}}$]{
    \includegraphics[width=7.2cm,height=5.9cm] {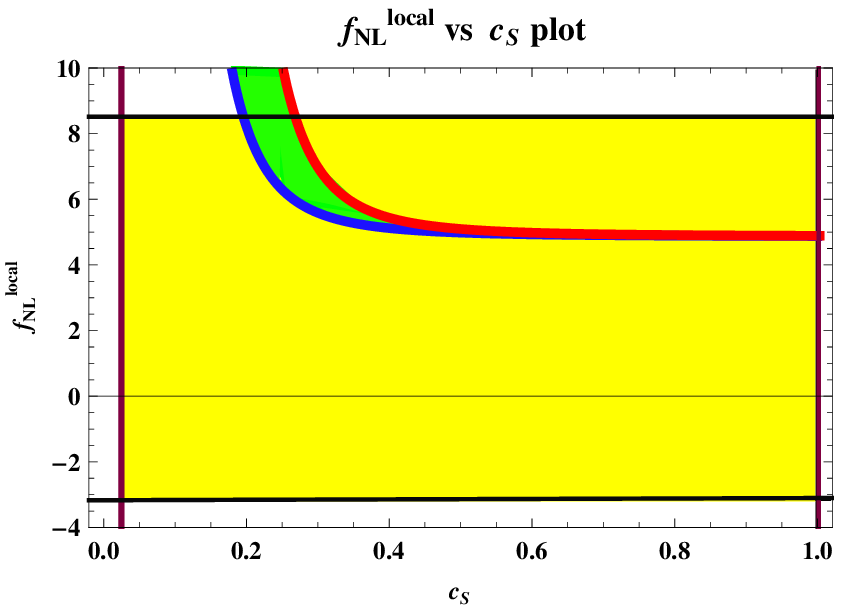}
    \label{fig:subfig1}
}
\subfigure[$\textcolor{blue}{\bf \underline{Case ~II}}$]{
    \includegraphics[width=7.2cm,height=5.9cm] {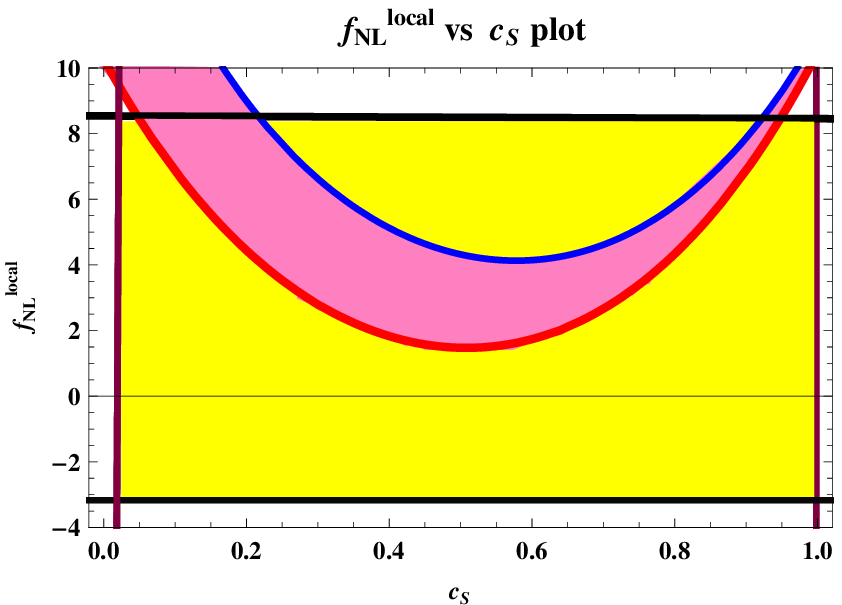}
    \label{fig:subfig2}
}
\subfigure[$\textcolor{blue}{\bf \underline{Case ~III}}$]{
    \includegraphics[width=7.2cm,height=5.9cm] {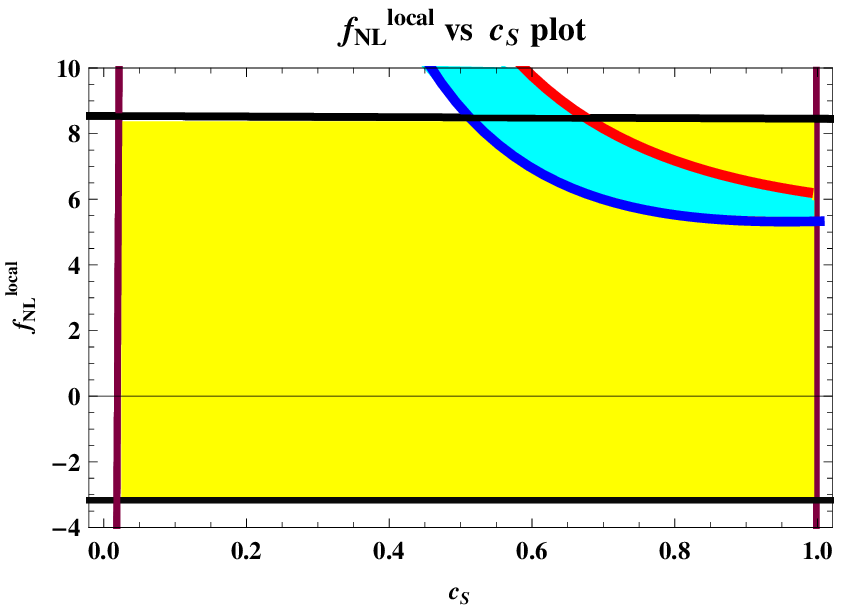}
    \label{fig:subfig3}
}
\subfigure[$\textcolor{blue}{\bf \underline{Case ~IV}}$]{
    \includegraphics[width=7.2cm,height=5.9cm] {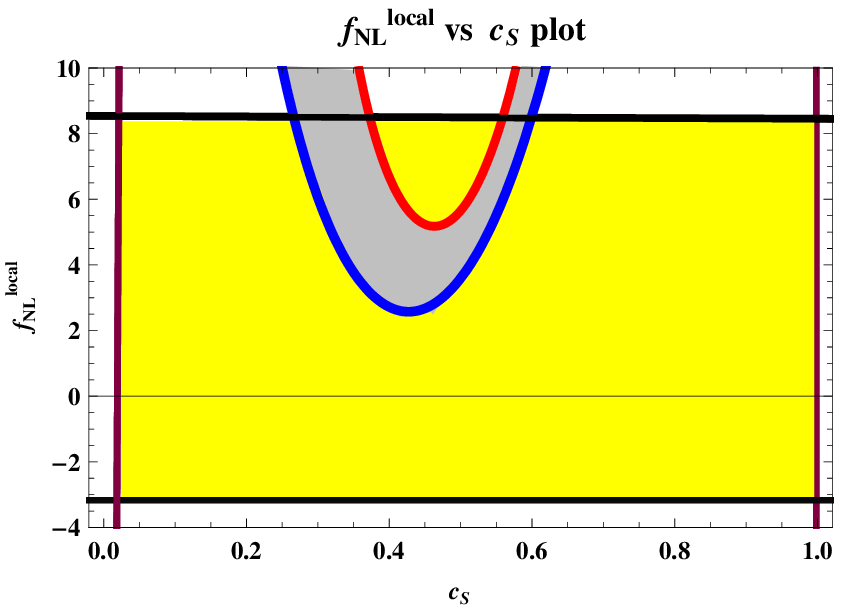}
    \label{fig:subfig4}
}
\caption[Optional caption for list of figures]{Behaviour
 of the local type of non-Gaussian parameter $f_{NL}^{\mathrm{local}}$ computed from the effective theory of ${\cal N}=1$ supergravity
with respect to the sound speed $c_{s}$ in the Hubble induced inflection point inflationary regime, represented
by $H>>m_{\phi}$. The shaded yellow region represents the allowed parameter space for Hubble induced inflation which satisfies 
the combined Planck constraints on the $f_{NL}^{\mathrm{local}}$ (within $1\sigma$ CL) and sound speed $c_{s}$ (within $2\sigma$ CL).
 The red, blue coloured boundaries and the bounded dark coloured regions are obtained from the scanning range of the 
scale of the of heavy scalar degrees freedom $M_{s}$ corresponds to the hidden sector,
 within the window $9.50\times 10^{10}~{\rm GeV}\,\leq M_s \leq 1.77\times10^{16}~{\rm GeV}\,$. The four distinctive
 features are obtained by varying the model parameters of the effective theory of ${\cal N}=1$ SUGRA, $c_{H},a_{H},M_{s}$ and $\phi_{0}$, 
subject to the constraint as stated in Eq~(\ref{P-space}-\ref{pj0}). The overlapping region between the dark coloured and yellow region 
satisfied the combined constraints on the $f_{NL}^{\mathrm{local}}$ and $c_{s}$ within our proposed framework and the rest of the regions are 
excluded.} 
\label{fza}
\end{figure*}

\begin{figure*}[htb]
\centering
\subfigure[$\textcolor{blue}{\bf \underline{Case ~I}}$]{
    \includegraphics[width=7.2cm,height=6.5cm] {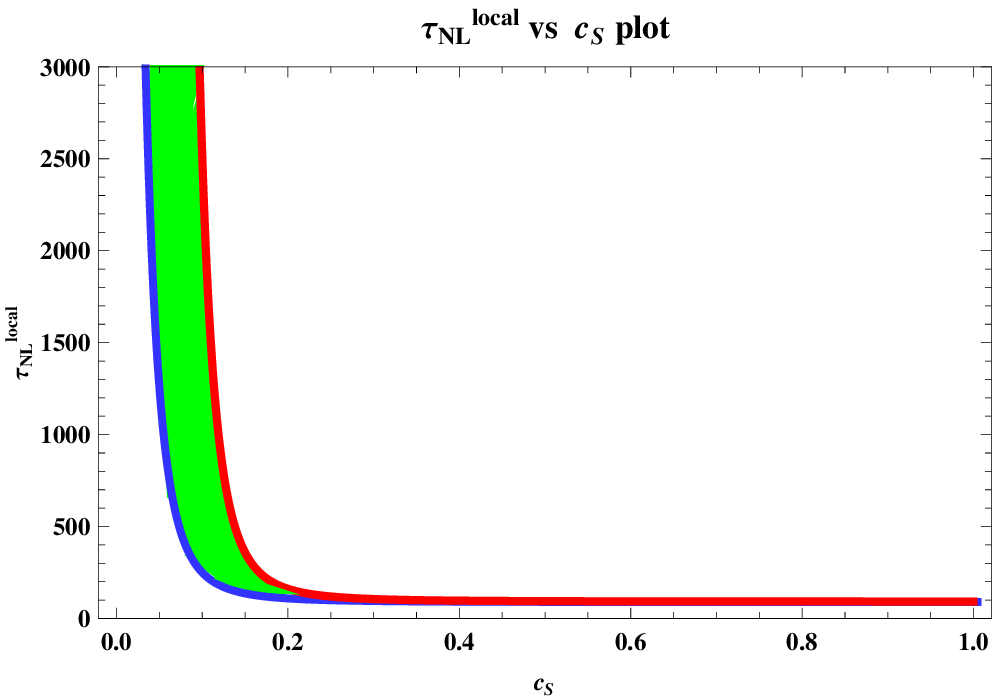}
    \label{fig:subfig1}
}
\subfigure[$\textcolor{blue}{\bf \underline{Case ~II}}$]{
    \includegraphics[width=7.2cm,height=6.5cm] {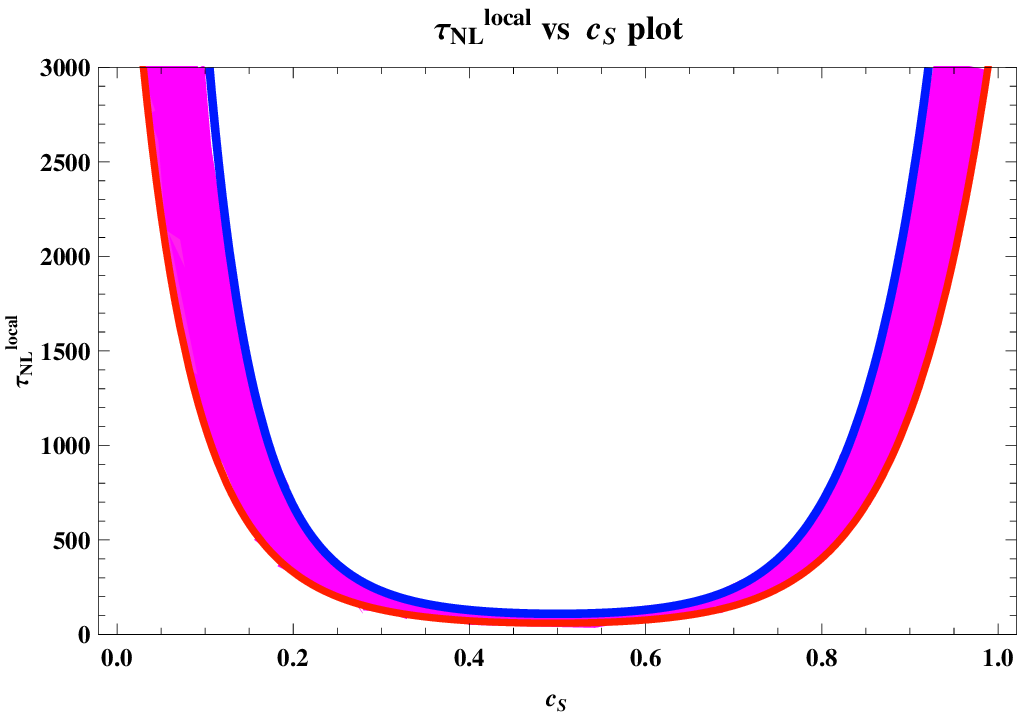}
    \label{fig:subfig2}
}
\subfigure[$\textcolor{blue}{\bf \underline{Case ~III}}$]{
    \includegraphics[width=7.2cm,height=6.5cm] {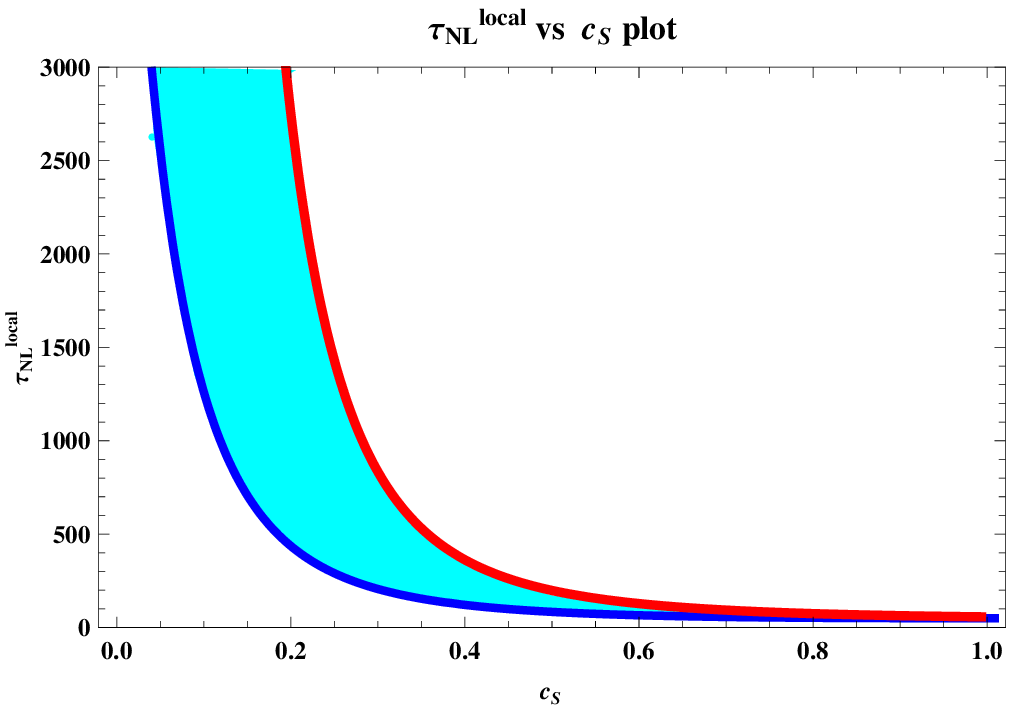}
    \label{fig:subfig3}
}
\subfigure[$\textcolor{blue}{\bf \underline{Case ~IV}}$]{
    \includegraphics[width=7.2cm,height=6.5cm] {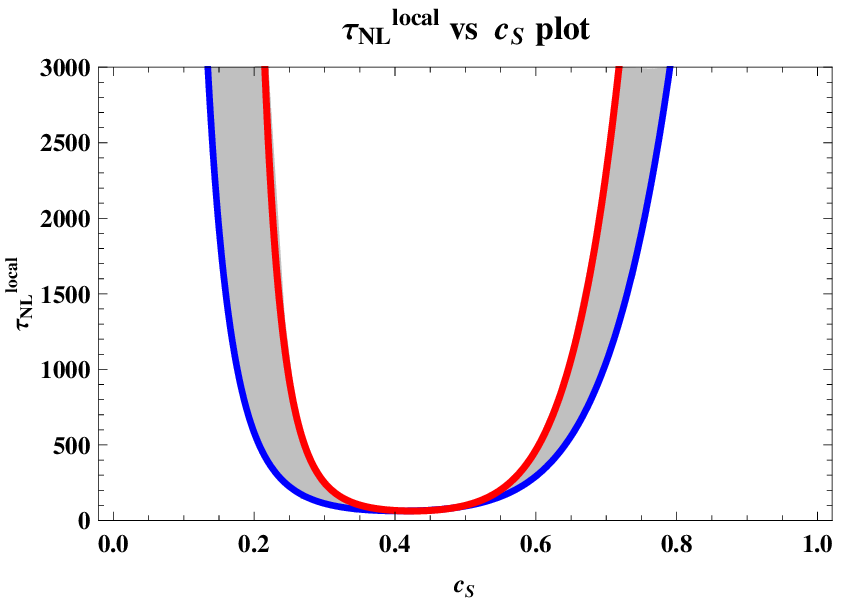}
    \label{fig:subfig4}
}
\caption[Optional caption for list of figures]{Behaviour
 of the local type of non-Gaussian parameter $\tau_{NL}^{\mathrm{local}}$ computed from the effective theory of ${\cal N}=1$ supergravity
with respect to the sound speed $c_{s}$ in the Hubble induced inflationary regime is represented
by $H>>m_{\phi}$. The red and blue coloured boundaries are obtained from the scanning range of the 
scale of the of heavy scalar degrees freedom $M_{s}$ corresponds to the hidden sector,
 within the window $9.50\times 10^{10}~{\rm GeV}\,\leq M_s \leq 1.77\times10^{16}~{\rm GeV}\,$. The four distinctive
 features are obtained by varying the model parameters of the effective theory of ${\cal N}=1$ SUGRA, $c_{H},a_{H},M_{s}$ and $\phi_{0}$, 
subject to the constraint as stated in Eq~(\ref{P-space}-\ref{pj0}). The dark coloured region 
satisfied the combined constraints on the $f_{NL}^{\mathrm{local}}$ and $c_{s}$ within the proposed framework. As Planck puts an upper bound,
$\tau_{NL}^{\mathrm{local}}\leq2800$, the rest of the region above the $\tau_{NL}^{\mathrm{local}}=2800$ line is 
excluded.} 
\label{fzat}
\end{figure*}

\begin{figure*}[htb]
\centering
\subfigure[$\textcolor{blue}{\bf \underline{Case ~I}}$]{
    \includegraphics[width=7.2cm,height=6.7cm] {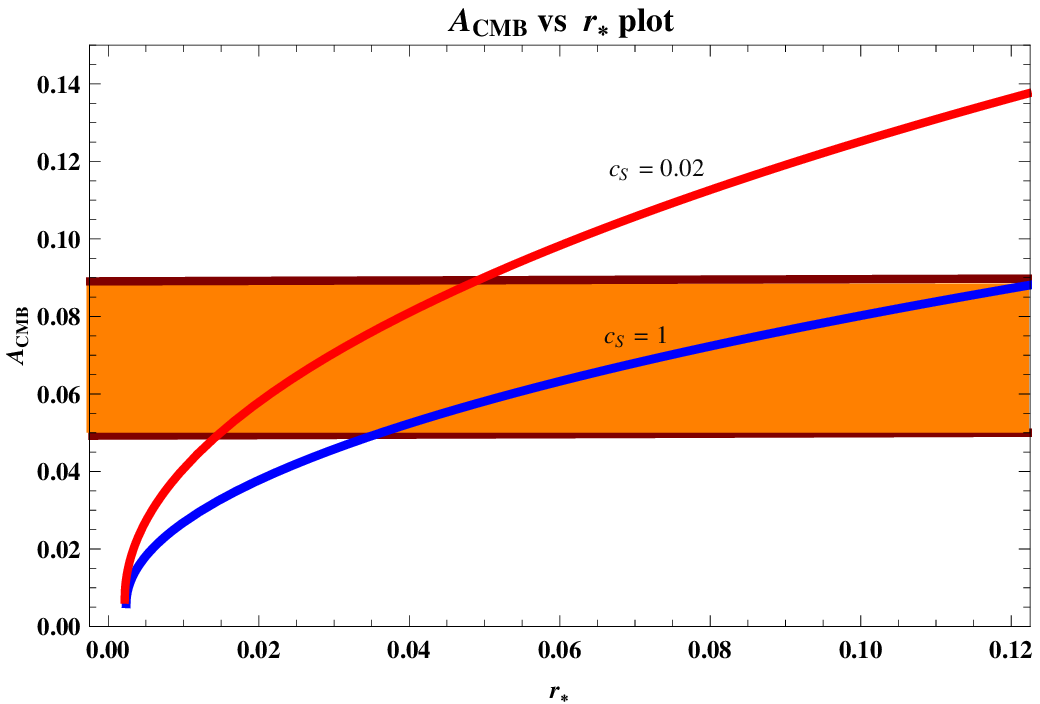}
    \label{fig:subfig1}
}
\subfigure[$\textcolor{blue}{\bf \underline{Case ~II}}$]{
    \includegraphics[width=7.2cm,height=6.7cm] {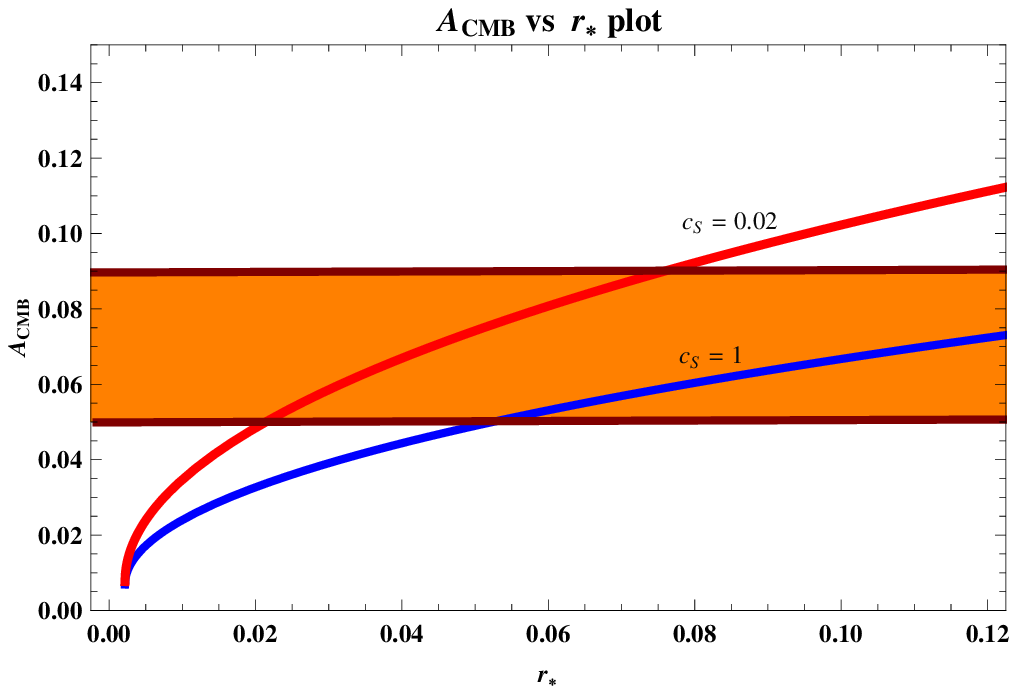}
    \label{fig:subfig2}
}
\subfigure[$\textcolor{blue}{\bf \underline{Case ~III}}$]{
    \includegraphics[width=7.2cm,height=6.7cm] {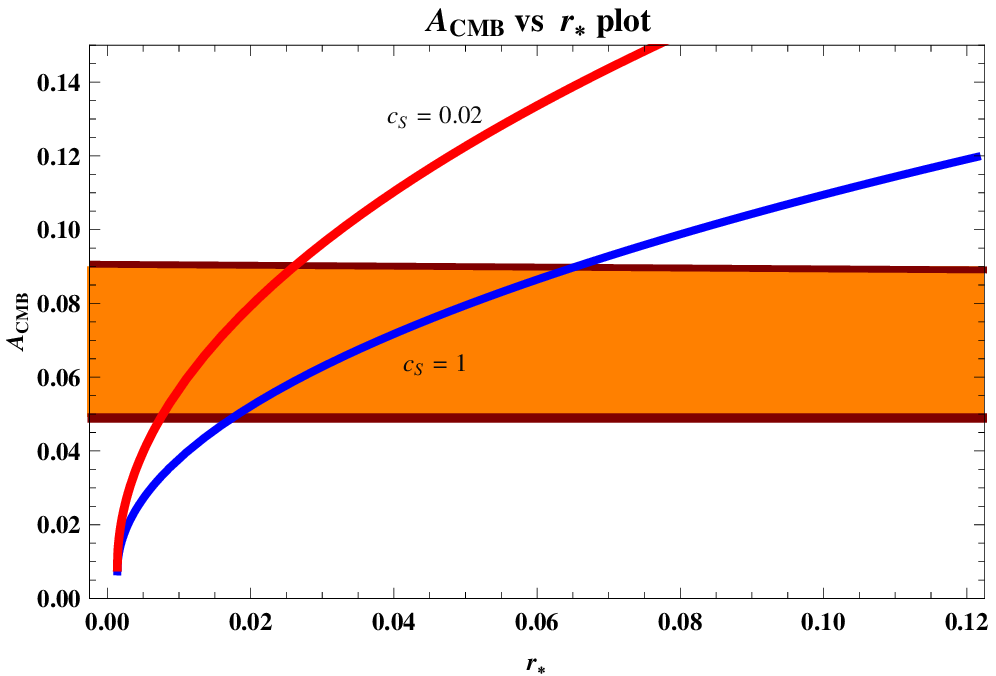}
    \label{fig:subfig3}
}
\subfigure[$\textcolor{blue}{\bf \underline{Case ~IV}}$]{
    \includegraphics[width=7.2cm,height=6.7cm] {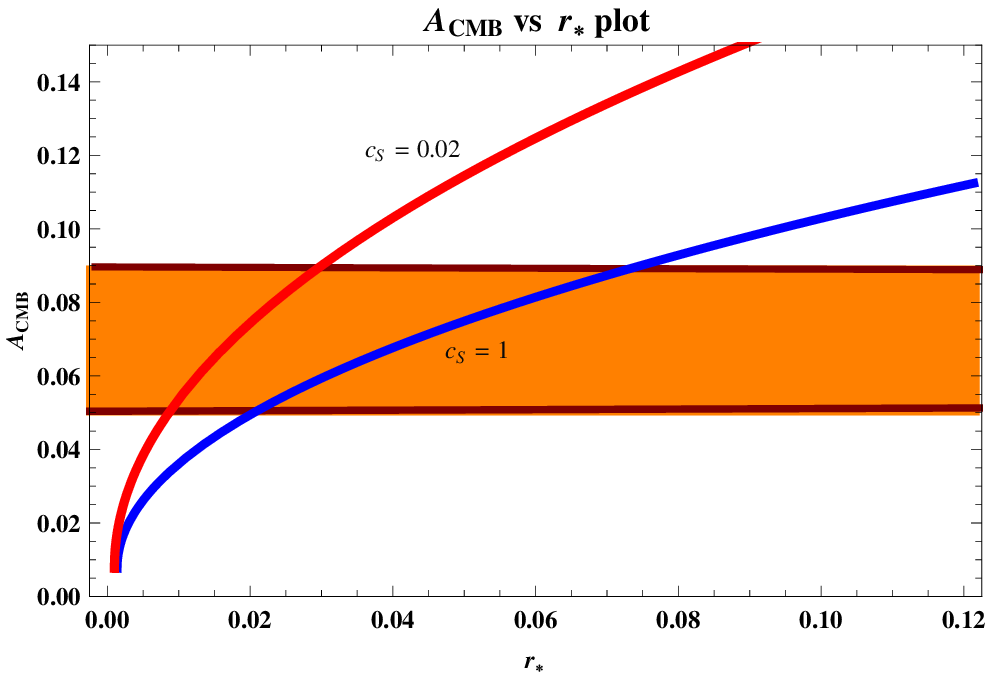}
    \label{fig:subfig4}
}
\caption[Optional caption for list of figures]{Behaviour
 of the CMB dipolar asymmetry parameter $A_{CMB}$ computed from the effective theory of ${\cal N}=1$ supergravity
with respect to the tensor-to-scalar ratio $r_{*}$ at the pivot scale, $k_{*}\sim 0.002~{\rm Mpc}^{-1}$ for the Hubble induced inflation.
 The red and blue coloured boundaries are obtained by fixing the sound speed at $c_{S}=0.02$ and $c_{S}=1$. The four distinctive
 features are obtained by varying the model parameters of the effective theory of ${\cal N}=1$ SUGRA, $c_{H},a_{H},M_{s}$ and $\phi_{0}$
subject to the constraint as stated in Eq~(\ref{P-space}-\ref{pj0}). The orange dark coloured region 
satisfied the Planck constraint on the $A_{CMB}$ within the proposed framework. Here only the region bounded by the red, blue and brown colour is the allowed one and
the rest of the region is 
excluded by the Planck data.} 
\label{fzata}
\end{figure*}

My present job is to now scan the parameter space for $c_H,~a_H$ with the help of,  by fixing $\lambda ={\cal O}(1)$ and $\delta \sim 10^{-4}$. 
In order to satisfy the inflationary paradigm, the Planck observational constraints, as stated in the introduction of the paper, I obtain the following
constraints on our parameters for $H_{inf}\geq m_{\phi}\sim {\cal O}(\rm TeV)$:
\begin{eqnarray}\label{P-space}
c_{H} &\sim &{\cal O}(10-10^{-6})\,,
\\
a_{H} & \sim &{\cal O}(30 - 10^{-3} )\,,
\\
M_s &\sim & {\cal O}(9.50\times 10^{10}-1.77\times10^{16})~{\rm GeV}\,,
\,.
\end{eqnarray}

Inflation would not occur outside the scanning region since, at least, one of the constraints would be violated.
Note that for the above ranges, the VEV of the inflaton, $\langle \phi \rangle =\phi_0$, gets automatically fixed by Eq.~(\ref{phi0}), 
in the sub-Planckian scale as:%
\begin{equation}\label{pj0}
\phi_{0} \sim {\cal O}(10^{14} -10^{17})~{\rm GeV}\,
\end{equation}
which bounds the tensor-to-scalar ratio within, $10^{-22}\leq r_{*}\leq0.12$ for the present setup.
This analysis will further constrain the non-minimal K\"ahler coupling parameters $a,b,c,d$~
\footnote{The analytical expressions for the non-minimal coupling parameters, $a,b,c,d$ can 
be expressed in terms of the scale (VEV) of the heavy field $M_{s}$ are explicitly mentioned in the appendix.}
appearing in the higher dimensional Planck scale suppressed 
opeartors within the following range:
\begin{eqnarray}\label{abcd-space}
a &\sim &{\cal O}(1-0.99)\,,
\\
b & \sim &{\cal O}(1-0.92)\,,
\\
c &\sim & {\cal O}(0.3-1)\,,
\\
d &\sim & {\cal O}(1-0.5)\,.
\end{eqnarray}

In Fig~(\ref{fza}) and Fig~(\ref{fzat}) I have shown the behaviour of the local type of non-Gaussian parameter $f_{NL}^{\mathrm{local}}$ and $\tau_{NL}^{\mathrm{local}}$
with respect to the sound speed $c_{s}$ in the Hubble induced inflationary regime ($H>>m_{\phi}$).
  In Fig~(\ref{fza}), the shaded yellow region represent the allowed parameter space for Hubble induced inflation which satisfies 
the combined Planck constraints on the $f_{NL}^{\mathrm{local}}$ and $c_{s}$. For all the four cases, the region above the $f_{NL}^{\mathrm{local}}=8.5$
line is observationally excluded by the Planck data. The four distinctive
 features are obtained by varying the model parameters of the effective theory of ${\cal N}=1$ SUGRA, $c_{H},a_{H}$ and $M_{s}$
subject to the constraint as stated in Eq~(\ref{P-space}-\ref{pj0}). As Planck puts an upper bound,
$\tau_{NL}^{\mathrm{local}}\leq2800$, the rest of the region above the $\tau_{NL}^{\mathrm{local}}=2800$ line in Fig~(\ref{fzat}) is 
excluded. In the present setup I have obtained the following stringent bound on
the $f_{NL}^{\mathrm{local}}$, $\tau_{NL}^{\mathrm{local}}$ and $g_{NL}^{\mathrm{local}}$ within the following range: 
\be\begin{array}{lll}\label{we1}
    \displaystyle 5\leq f_{NL}^{\mathrm{local}}\leq8.5,~~~100\leq\tau_{NL}^{\mathrm{local}}\leq2800,~~~23.2\leq g_{NL}^{\mathrm{local}}\leq648.2 & \mbox{ for $\underline{\bf Case ~I}$}  \\ 
  1\leq f_{NL}^{\mathrm{local}}\leq8.5,~~~150\leq\tau_{NL}^{\mathrm{local}}\leq2800,~~~34.7\leq g_{NL}^{\mathrm{local}}\leq648.2 & \mbox{ for $\underline{\bf Case ~II}$}  \\ 
    5\leq f_{NL}^{\mathrm{local}}\leq8.5,~~~~75\leq\tau_{NL}^{\mathrm{local}}\leq2800,~~~17.4\leq g_{NL}^{\mathrm{local}}\leq648.2 & \mbox{ for $\underline{\bf Case ~III}$}  \\ 
    2\leq f_{NL}^{\mathrm{local}}\leq8.5,~~~110\leq\tau_{NL}^{\mathrm{local}}\leq2800,~~~25.5\leq g_{NL}^{\mathrm{local}}\leq648.2 & \mbox{ for $\underline{\bf Case ~IV}$}.
          \end{array}
\ee 
Here the theoretical upper and lower bound on $f_{NL}^{\mathrm{local}}$~\footnote{In the prescribed setup the consistency relation between the
non-Gaussian parameter $f_{NL}^{\mathrm{local}}$ and the spectral tilt $n_{s}$ \cite{Maldacena:2002vr}, $f_{NL}^{\mathrm{local}}\sim\frac{5}{12}(1-n_{s})$, does not hold as in the present 
setup sound speed, $c_{s}\neq 1$ and for such non-minimal ${\cal N}=1$ SUGRA setup, Planck data favours lower values of the sound speed (within $0.02<c_{s}<1$).}
 satisfy both the constraints on the $f_{NL}^{\mathrm{local}}$ and $c_{s}$
observed by Planck data. Also it is important to note that, within this prescribed framework, $\tau_{NL}^{\mathrm{local}}$ is bounded by below
for all the four cases and consequently it is possible to put a stringent lower bound on $\tau_{NL}^{\mathrm{local}}$ which satisfies 
the constraints on $\tau_{NL}^{\mathrm{local}}$ and $c_{s}$ both. Till date the observational results obtained from Planck do not give any significant 
constraint on $g_{NL}^{\mathrm{local}}$. However in this paper I have provided a theoretical lower and upper bound of $g_{NL}^{\mathrm{local}}$ using the 
consistency relation between $\tau_{NL}^{\mathrm{local}}$ and $g_{NL}^{\mathrm{local}}$ as stated in Eq~(\ref{conq}).

Finally, in Fig~(\ref{fzata}) I have shown the behaviour
 of the CMB dipolar asymmetry parameter $A_{CMB}$ 
with respect to the tensor-to-scalar ratio $r_{*}$ within, $10^{-22}\leq r_{*}\leq0.12$, at the pivot scale, $k_{*}\sim 0.002~{\rm Mpc}^{-1}$ for the Hubble induced inflation.
 Here the red and blue coloured boundaries are obtained by fixing the sound speed at $c_{S}=0.02$ and $c_{S}=1$. The orange dark coloured region 
satisfied the Planck constraint
 on the $A_{CMB}$ i.e. $0.05\leq A_{CMB}\leq 0.09$~\footnote{The upper bound of the CMB dipolar asymmetry parameter ($A_{CMB}$) can be expressed in terms of the
non-Gaussian parameter $f_{NL}^{\mathrm{local}}$ through a consistency relation as \cite{Namjoo:2013fka}, $A_{CMB}\lesssim 10^{-1}f_{NL}^{\mathrm{local}}$, which perfectly holds good
in the present effective theory setup.} 
for $10^{-22}\leq r_{*}\leq0.12$ within our proposed framework. In Fig~(\ref{fzata}) 
only the region bounded by the red, blue and brown colour is the allowed one and
the rest of the region ($A_{CMB}<0.02$ and $A_{CMB}>0.09$) is 
excluded by the Planck data.

\section{Conclusion}

In this paper, I have shown that in any general class of ${\cal N}=1$
SUGRA inflationary framework, the behaviour of K\"ahler potential in presence of non-minimal K\"ahler corrections in effective theory setup 
are constrained via the non-minimal couplings of the non-renormalizable gauge invariant K\"ahler higher dimensional Planck scale suppressed 
operators from the observational constraint on non-Gaussianity, sound speed and CMB dipolar asymmetry as obtained from the Planck data.
In the present setup the hidden sector based heavy field is settled down in its potential via its Hubble induced vacuum energy density. 
In particular, for the numerical estimations in this paper I have used a very particular kind of (inflection point) inflationary model, which 
is fully embedded within MSSM, where the inflaton is made up of $\widetilde L\widetilde L\widetilde e$ and $\widetilde u\widetilde d\widetilde d$
gauge invariant D-flat directions. However the prescribed methodology holds good for other kinds of inflationary models too. 
 
Further I have scanned the multiparameter region characterized by the Hubble induced mass parameter, $c_{H}$, A-term, $a_{H}$ and the scale of the 
heavy field $M_{s}$, where I have satisfied the current Planck observational constraints on the, 
inflationary parameters: $P_S,~n_S,~c_s,r_*$ (within $2\sigma$~CL), non-Gaussian parameters: $f_{NL}^{local},\tau_{NL}^{local}$ 
(within $1\sigma-1.5\sigma$~CL) and CMB dipolar asymmetry parameter $A_{CMB}$ (within $2\sigma$~CL).
Consequently the non-minimal K\"ahler couplings, $a,b,c,d$ are fixed within $\sim {\cal O}(1)$ in the present effective theory setup.
Finally, using this methodology, I have obtained 
the theoretical upper and lower bound on the non-Gaussian parameters within the range, ${\cal O}(1-5)\leq f_{NL}\leq8.5$, ${\cal O}(75-150)\leq\tau_{NL}<2800$ 
and ${\cal O}(17.4-34.7)\leq g_{NL}\leq 648.2$, and the
 CMB dipolar asymmetry parameter within, $0.05\leq A_{CMB}\leq0.09$, which satisfy 
the observational constraints stated in Eq~(\ref{pow}-\ref{f-nl3}), as obtained from Planck data. 

There is also a future prospect of upgrading the present methodology proposed in this paper by studying the further stringent phenomenological 
constraints on the non-minimal couplings $a,b,c,d$, as appearing in the context of
higher dimensional Planck scale suppressed K\"ahler operators within ${\cal N}=1$ SUGRA
 by imposing
 the constraint on Higgs mass \cite{Aad:2012tfa,Chatrchyan:2013lba} and the dark matter abundance \cite{Ade:2013zuv,Boehm:2012rh}
 via WIMPy baryogenesis scenario (see also Refs.~\cite{Cui:2011ab,Bernal:2012gv,Bernal:2013bga} for
 the various theoretical issues). A detailed analysis on these aspects have been reported shortly as a separate paper \cite{Sayan:2014}.

\section*{Acknowledgments}
I would like to thank Prof. Anupam Mazumadar for the huge support and helpful suggestions throughout the project.  
I am indebted to Prof. Daniel Baumann for the various useful discussions. 
I also thank Council of Scientific and
Industrial Research, India for financial support through Senior
Research Fellowship (Grant No. 09/093(0132)/2010). 
 I take this opportunity to thank the organizers of 8th Asian School on Strings, Particles and Cosmology, 2014 for the hospitality during the
work. Last but not the least I would like to thank sincerely to Prof. Soumitra SenGupta and Dr. Supratik Pal for their constant support and inspiration.  

\section*{Appendix}

  
\subsection*{A. The model parameters $\alpha,\beta,\gamma,\kappa$:}
The model parameters characterizing the potential stated in Eq~(\ref{rt1a}) can be expressed as:  
\begin{eqnarray}\label{p1}
     \alpha&=&M^{4}_{s}+\left(\frac{(n-2)^{2}}{n(n-1)}+\frac{(n-2)^2}{n}\delta^{2}\right)c_{H}H^{2}\phi^{2}_{0}+\cdots,\\
 \beta&=&2\left(\frac{n-2}{2}\right)^{2}\delta^{2}c_{H}H^{2}\phi_{0}+\cdots,\\
 \gamma&=&\frac{c_{H}H^{2}}{\phi_{0}}\left(4(n-2)^2-\frac{(n-1)(n-2)^3}{2}\delta^{2}\right)+\cdots,\\
 \small\kappa&=&\frac{c_{H}H^{2}}{\phi^{2}_{0}}\left(12(n-2)^3-\frac{(n-1)(n-2)(n-3)
(7n^2-27n+26)}{2}\delta^{2}\right)+\cdots
   \end{eqnarray}
where the higher order $\cdots$ terms are neglected due to $\delta^{2}<<1$. During numerical estimations I fix $n=6$ for
 $\widetilde L\widetilde L\widetilde e$ and $\widetilde u\widetilde d\widetilde d$  D-flat 
directions respectively.

\subsection*{B. The symbol $\Sigma=X,Y,Z,W$:}

The symbols appearing in the Eq~(\ref{sigbz}), in the definition of the sound speed $c_s$ for $s<< M_{p}$, after imposing the slow-roll approxiation are given by:
 \be\begin{array}{lll}\label{de23}
 \displaystyle {\bf X}_{1}(t)= \sqrt{\frac{2\epsilon_{V}(\phi)V(\phi)}{3}}
\left\{ \sqrt{\frac{2\epsilon_{V}(\phi)V(\phi)}{3}}\frac{aM^{3}_{s}}{M^{2}_{p}}\left[2\sin(2M_{s}t)+4\cos(M_{s}t)\right]\right.\\ \left.
\displaystyle~~~~~~~~~~~~~~~~~~~~~~~~~~~~~~~~~~~~~~~~~~~~
-\frac{aM^{4}_{s}}{M^{2}_{p}}\lvert\phi\rvert\cos{\bf\Theta}\left[\cos(2M_{s}t)-\sin(M_{s}t)\right]\right\},\end{array}\ee
\be\begin{array}{lll}\label{de23a}\displaystyle {\bf Y}_{1}(t)= \sqrt{\frac{2\epsilon_{V}(\phi)V(\phi)}{3}}\left\{ \sqrt{\frac{2\epsilon_{V}(\phi)V(\phi)}{3}}\frac{2bM^{2}_{s}}{M_{p}}\cos(M_{s}t)
+\frac{bM^{3}_{s}}{M_{p}}\lvert\phi\rvert\cos{\bf\Theta}\sin(M_{s}t)\right\},\end{array}\ee
\be\begin{array}{lll}\label{de23aa}\displaystyle {\bf Z}_{1}(t)= \sqrt{\frac{2\epsilon_{V}(\phi)V(\phi)}{3}}\left\{\sqrt{\frac{2\epsilon_{V}(\phi)V(\phi)}{3}}\frac{cM^{3}_{s}}{4M^{2}_{p}}\left[2\sin(2M_{s}t)+4\cos(M_{s}t)\right]
\right.\\ \left.
\displaystyle~~~~~~~~~~~~~~~~~~~~~~~~~~~~~~~~~~~~~~~~~~~
-\frac{cM^{4}_{s}}{4M^{2}_{p}}\lvert\phi\rvert\cos{\bf\Theta}\left[\cos(2M_{s}t)-\sin(M_{s}t)\right]\right\},\end{array}\ee
\be\begin{array}{lll}\label{de23aaa}\displaystyle {\bf W}_{1}(t)= \sqrt{\frac{2\epsilon_{V}(\phi)V(\phi)}{3}}\left\{ \sqrt{\frac{2\epsilon_{V}(\phi)V(\phi)}{3}}\frac{4dM^{2}_{s}}{M_{p}}\cos(M_{s}t)
+\frac{dM^{3}_{s}}{M_{p}}\lvert\phi\rvert\cos{\bf\Theta}\sin(M_{s}t)\right\},\end{array}\ee
\be\begin{array}{lll}\label{de23aaaa}\displaystyle {\bf X}_{2}(t)=\left({\bf Y}_{2}(t)+\frac{a|\phi|^{2}M^{5}_{s}}{M^{2}_{p}}\sin(2M_{s}t)\right),\\
\displaystyle {\bf Y}_{2}(t)={\bf Z}_{2}(t)={\bf W}_{2}(t)=5M^{5}_{s}\sin(2M_{s}t)+8M^{5}_{s}\cos(M_{s}t),\\
\displaystyle {\bf X}_{3}(t)=\left({\bf Y}_{3}(t)-\frac{a|\phi|^{2}M^{5}_{s}}{M^{2}_{p}}\sin(2M_{s}t)\right),\\
\displaystyle {\bf Y}_{3}(t)={\bf Z}_{3}(t)={\bf W}_{3}(t)=3M^{5}_{s}\sin(2M_{s}t)-8M^{5}_{s}\cos(M_{s}t).\\
    \end{array}\ee
Here the complex inflaton field $\phi$ is parameterized by, $\phi=\lvert\phi\rvert\exp(i{\bf\Theta})$. Here the new parameter ${\bf\Theta}$ characterizes 
the phase factor associated with the inflaton and it has a two dimensional rotational symmetry.
\subsection*{C. Expression for the non-minimal couplings $a,b,c,d$:}\label{abcd}
The expressions for the non-minimal supergravity coupling parameter $a,b,c,$ and $d$ for all the four physical cases within ${\cal N}=1$ SUGRA   
 with $H_{inf}>>m_{\phi}$ can be expressed in terms of the VEV of the heavy field, $\langle s \rangle= M_{s}$ as:
\be\begin{array}{lll}\label{pspace1cv}
    \displaystyle a\sim {\cal O}\left(1-1.06\times 10^{-5}\frac{n^{2}}{(n-1)}\frac{M^{2}_{s}}{M^{2}_{p}}\right)& \mbox{ for $\underline{\bf Case ~I}$},\\
\displaystyle b\sim {\cal O}\left(\sqrt{\left|\frac{\left(3-\frac{1}{n}\right)^{2}}{100(n-1)}\frac{M^{2}_{s}}{M^{2}_{p}}-1\right|}\right)& \mbox{ for $\underline{\bf Case ~II}$},\\
\displaystyle c\sim {\cal O}\left(\frac{1}{500}\left|\frac{\pm 8.16\frac{M_{s}}{M_{p}}\sqrt{n-1}-\sqrt[4]{\frac{2}{3}}\left(\sqrt{\frac{3}{2}}(n-3)+1
\right)}{1.24\frac{M_{s}}{M_{p}}-\sqrt[4]{\frac{2}{3}}n}\right|\right)& \mbox{ for $\underline{\bf Case ~III}$},\\
\displaystyle d\sim {\cal O}\left(\sqrt{\left|2.54\times 10^{-4}\frac{\left(n-1+\sqrt{6}\right)^{2}}{(n-1)}\frac{M^{2}_{s}}{M^{2}_{p}}-1\right|}\right)& \mbox{ for $\underline{\bf Case ~IV}$}.
   \end{array}\ee


\subsection*{D. Expression for $\Sigma_{s}(t),\Xi_{s}(t),\Psi_{s}(t),\Theta_{s}(t)$:}\label{time dep}

\be\begin{array}{lll}\label{sig1}
    \tiny  \Sigma_{s}(t)=
\frac{e^{\left(-3H-\sqrt{\frac{4aM^{4}_{s}}{M^{2}_{p}\left(1+
\frac{aM^{2}_{s}}{4M^{2}_{p}}\right)}+9H^{2}}\right)t}}{
\frac{3aM^{4}_{s}}{M^{2}_{p}\left(1+\frac{aM^{2}_{s}}{4M^{2}_{p}}\right)}\left(
\frac{aM^{4}_{s}}{M^{2}_{p}\left(1+\frac{aM^{2}_{s}}{4M^{2}_{p}}\right)}+2H^{2}\right)}\left\{
\frac{3\gamma}{\left(1+\frac{aM^{2}_{s}}{4M^{2}_{p}}\right)}\left[\left(\frac{aM^{4}_{s}}{M^{2}_{p}\left(1+
\frac{aM^{2}_{s}}{4M^{2}_{p}}\right)}+3H^{2}\right.\right.\right.\\ \left.\left.\left. 
~~~~~~~~~~~~~~~~~~~\tiny-H\sqrt{\frac{4aM^{4}_{s}}{M^{2}_{p}\left(1+\frac{aM^{2}_{s}}{4M^{2}_{p}}\right)}+9H^{2}}\right)
\left({\bf C}^{2}_{1}+{\bf C}^{2}_{2}e^{2\left(
\sqrt{\frac{4aM^{4}_{s}}{M^{2}_{p}\left(1+\frac{aM^{2}_{s}}{4M^{2}_{p}}\right)}+9H^{2}}\right)t}\right)\right.\right. \\ \left.\left.~~~~~~~~~~~~~~~~~~~~~~~~~~
~~~~~~~~~~~~~~~~~~~~~~~~~~~~~~~~~~\tiny-6{\bf C}_{1}{\bf C}_{2}\left(
\frac{aM^{4}_{s}}{M^{2}_{p}\left(1+\frac{aM^{2}_{s}}{4M^{2}_{p}}\right)}+2H^{2}\right)\right]\right\},\end{array}\ee
\be\begin{array}{lll}\label{sig2}
\tiny \Xi_{s}(t)=\gamma\left[\frac{{\bf C}^{2}_{4}}{54H^{4}}e^{-6Ht}+
\frac{t}{81H^{5}}\left\{\beta^{2}\left(2-3Ht\right)
-3H^{2}\beta\left(9Ht{\bf C}_{3}
\right.\right.\right.\\ \left.\left.\left.~~~~~~~~~~~~~~~~~~~~~~~~
-\left[\beta t^{2}+6{\bf C}_{3}\right]\right)-81H^{4}
\left(\frac{\beta}{\gamma}-{\bf C}^{2}_{3}\right)\right\}-\frac{{\bf C}_{4}}{243H^{5}}e^{-3Ht}\right.\\ \left.
~~~~~~~~~~~~~~~~~~~~~~~~~~~~~~~~~~~~~~~~~\times\left\{\beta
 \left(2 + 6 H t + 9 H^2 t^2\right) - 18 H^{2}\left(1 + 3 H t\right) {\bf C}_{3}\right\}\right],\end{array}\ee
\be\begin{array}{lll}\label{sig3}
\tiny \Psi_{s}(t)=\gamma\left[\frac{{\bf C}^{2}_{6}}{54H^{4}}e^{-6Ht}+
\frac{t}{81H^{5}}\left\{\beta^{2}\left(2-3Ht\right)
-3H^{2}\beta\left(9Ht{\bf C}_{5}
\right.\right.\right.\\ \left.\left.\left.~~~~~~~~~~~~~~~~~~~~~~~~
-\left[\beta t^{2}+6{\bf C}_{5}\right]\right)-81H^{4}
\left(\frac{\beta}{\gamma}-{\bf C}^{2}_{5}\right)\right\}-\frac{{\bf C}_{6}}{243H^{5}}e^{-3Ht}\right.\\ \left.
~~~~~~~~~~~~~~~~~~~~~~~~~~~~~~~~~~~~~~~~~\times\left\{\beta
 \left(2 + 6 H t + 9 H^2 t^2\right) - 18 H^{2}\left(1 + 3 H t\right) {\bf C}_{5}\right\}\right],\end{array}\ee
\be\begin{array}{lll}\label{sig4}
\tiny \Theta_{s}(t)=\frac{\gamma}{\left(1+\frac{2dM_{s}}{M_{p}}\right)}\left[\frac{{\bf C}^{2}_{8}}{54H^{4}}e^{-6Ht}+
\frac{t}{81H^{5}}\left\{\frac{\beta^{2}}{\left(1+\frac{2dM_{s}}{M_{p}}\right)^{2}}\left(2-3Ht\right)
-\frac{3H^{2}\beta}{\left(1+\frac{2dM_{s}}{M_{p}}\right)}\left(9Ht{\bf C}_{7}
\right.\right.\right.\\ \left.\left.\left.~~~~~~~~~~~~~~~~~~~~~~~~
-\left[\frac{\beta t^{2}}{\left(1+\frac{2dM_{s}}{M_{p}}\right)}+6{\bf C}_{7}\right]\right)-81H^{4}
\left(\frac{\beta}{\gamma}-{\bf C}^{2}_{7}\right)\right\}-\frac{{\bf C}_{8}}{243H^{5}}e^{-3Ht}\right.\\ \left.
~~~~~~~~~~~~~~~~~~~~~~~~~~~~~~~~~~~~~~~~~\times\left\{\frac{\beta}{\left(1+\frac{2dM_{s}}{M_{p}}\right)} 
\left(2 + 6 H t + 9 H^2 t^2\right) - 18 H^{2}\left(1 + 3 H t\right) {\bf C}_{7}\right\}\right].
   \end{array}\ee





\end{document}